\documentclass[article,apl,superscriptaddress,amsmath,amssymb,floatfix,twocolumn,tightenlines]{revtex4-2}

\usepackage{textcomp}
\usepackage{setspace}
\usepackage{amsmath}
\usepackage[margin=0.75in]{geometry}
\usepackage{physics}
\usepackage{amsfonts}
\usepackage{amssymb}
\usepackage{bm}
\usepackage{MnSymbol}
\usepackage{graphicx}
\usepackage{natbib}
\usepackage{gensymb}
\usepackage{setspace}
\usepackage{bbm}
\usepackage{tabularray}
\UseTblrLibrary{booktabs}

\usepackage{xspace}
\usepackage{graphicx}
\usepackage[dvipsnames]{xcolor}
\usepackage[separate-uncertainty=true,multi-part-units=single,retain-unity-mantissa=false]{siunitx}
\usepackage[version=4]{mhchem}
\usepackage[colorlinks, citecolor={blue}, linkcolor={black}, urlcolor={blue}]{hyperref}
\usepackage{xfp} 
\newcommand\ExtendedData{
    \xdef\preextfigures{\arabic{figure}}
    \renewcommand\thefigure{\textbf{\fpeval{\arabic{figure}-\preextfigures}}}
    \renewcommand{\figurename}{\textbf{\extlbl}}
    
    \xdef\preexttables{\arabic{table}}
    \renewcommand\thetable{\textbf{\fpeval{\arabic{table}-\preexttables}}}
    \renewcommand{\tablename}{\textbf{\exttbllbl}}
}

\newcommand{\figlbl}{Fig.}
\newcommand{\extlbl}{Extended~Data~Fig.} 
\newcommand{\extlbls}{Extended~Data~Figs.} 
\newcommand{\exttbllbl}{Extended~Data~Table}

\renewcommand{\figurename}{\textbf{\figlbl}} 
\renewcommand{\thefigure}{\textbf{\arabic{figure}}}
\newcommand{\figtitle}[1]{\textbf{#1}\xspace} 

\newcommand{\panel}[1]{\textbf{#1}\xspace}
\newcommand{\moire}{moir{\'e}\xspace}
\newcommand{\Moire}{Moir{\'e}\xspace}
\newcommand{\supermoire}{supermoir{\'e}\xspace}
\newcommand{\hb}{$\overline{\text{h}}$}
\newcommand{\rr}[1]{\mathrm{#1}}
\newcommand{\bu}{\vec{u}}

\newcommand{\misfitr}[3]{ V^{\mathrm{GSFE}}_{#1}(\vec{B}^{#2 \rightarrow #3})}

\begin{document}

\title{Helical trilayer graphene: a \moire platform for strongly-interacting topological bands}

\date{{\small \today}}

\author{Li-Qiao Xia}
\thanks{These authors contributed equally.}
\affiliation{Department of Physics, Massachusetts Institute of Technology, Cambridge, MA, USA}
\author{Sergio C. de la Barrera}
\thanks{These authors contributed equally.}
\affiliation{Department of Physics, Massachusetts Institute of Technology, Cambridge, MA, USA}
\affiliation{Department of Physics, University of Toronto, Toronto, Ontario, Canada}
\author{Aviram Uri}
\thanks{These authors contributed equally.}
\affiliation{Department of Physics, Massachusetts Institute of Technology, Cambridge, MA, USA}
\author{Aaron Sharpe}
\affiliation{Materials Physics Department, Sandia National Laboratories, Livermore, CA, USA}
\author{Yves H. Kwan}
\affiliation{Princeton Center for Theoretical Science, Princeton University, Princeton NJ 08544, USA}
\author{Ziyan Zhu}
\affiliation{Stanford Institute for Materials and Energy Sciences, SLAC National Accelerator Laboratory, Menlo Park, CA 94025}
\author{Kenji Watanabe}
\affiliation{Research Center for Electronic and Optical Materials, National Institute for Materials Science, 1-1 Namiki, Tsukuba 305-0044, Japan}
\author{Takashi Taniguchi}
\affiliation{Research Center for Materials Nanoarchitectonics, National Institute for Materials Science,  1-1 Namiki, Tsukuba 305-0044, Japan}
\author{David Goldhaber-Gordon}
\affiliation{Department of Physics, Stanford University, Stanford, CA 94305, USA}
\affiliation{Stanford Institute for Materials and Energy Sciences, SLAC National Accelerator Laboratory, Menlo Park, CA 94025}
\author{Liang Fu}
\affiliation{Department of Physics, Massachusetts Institute of Technology, Cambridge, MA, USA}
\author{Trithep Devakul}
\affiliation{Department of Physics, Stanford University, Stanford, CA 94305, USA}
\affiliation{Department of Physics, Massachusetts Institute of Technology, Cambridge, MA, USA}
\author{Pablo Jarillo-Herrero}
\affiliation{Department of Physics, Massachusetts Institute of Technology, Cambridge, MA, USA}

\begin{abstract}
Quantum geometry of electronic wavefunctions results in fascinating topological phenomena.
A prominent example is the intrinsic anomalous Hall effect (AHE) in which a Hall voltage arises in the absence of an applied magnetic field.
The AHE requires a coexistence of Berry curvature and spontaneous time-reversal symmetry breaking \cite{Nagosa2010Anomalous}. 
These conditions can be realized in two-dimensional \moire systems with broken $xy$-inversion symmetry ($C_{2z}$) that host flat electronic bands \cite{Sharpe2019Emergent, Serlin2020Intrinsic, Chen2020Tunable, Polshyn2020electrical, Chen2021Electrically, Tschirhart2021Imaging, He2021competing, Stepanov2021Competing, Polshyn2022Topological, Zhang2023local, Tseng2022anomalous, Kuiri2022, he2021chiralitydependent, zhang2023valley, li2021quantum, Lin2022spinorbit, polski2022hierarchy, Chen2022tunable, anderson2023programming, Cai2023Signatures, zeng2023integer, foutty2023mapping, park2023observation, xu2023observation}.
Here, we explore helical trilayer graphene (HTG) \cite{mora2019flatbands,mao2023supermoire,popov2023magic,devakul2023magicangle,guerci2023chern,nakatsuji2023multiscale,guerci2023natureofevenodd,kwan2023strongcoupling}, three graphene layers twisted sequentially by the same angle forming two misoriented \moire patterns.
Although HTG is globally $C_{2z}$-symmetric, surprisingly we observe clear signatures of topological bands.
At a magic angle $\theta_\mathrm{m}\approx\SI{1.8}{\degree}$, we uncover a robust phase diagram of correlated and magnetic states using magnetotransport measurements.
Lattice relaxation leads to large periodic domains in which $C_{2z}$ is broken on the \moire scale.
Each domain harbors flat topological bands with valley-contrasting Chern numbers $\pm(1,-2)$ \cite{devakul2023magicangle, guerci2023chern, nakatsuji2023multiscale, kwan2023strongcoupling}.
We find correlated states at integer electron fillings per \moire unit cell $\nu=1,2,3$ and fractional fillings $2/3,7/2$ with the AHE arising at $\nu=1,3$ and $2/3,7/2$.
At $\nu=1$, a time-reversal symmetric phase appears beyond a critical electric displacement field, indicating a topological phase transition.
Finally, hysteresis upon sweeping $\nu$ points to first-order phase transitions across a spatial mosaic of Chern domains \cite{Grover2022} separated by a network of topological gapless edge states.
We establish HTG as an important platform that realizes ideal conditions for exploring strongly interacting topological phases and, due to its emergent \moire-scale symmetries, demonstrates a novel way to engineer topology. 
\end{abstract}

\maketitle
\newpage

\section*{Introduction}
The combination of strong electronic correlations and non-trivial band topology is fertile ground for exotic electronic phenomena.
Driven by Berry curvature and orbital magnetization, the spontaneous emergence of the AHE in non-magnetic materials is a notable example.
It requires a periodic system with broken $PT$ and time-reversal ($T$) symmetries \cite{Xiao2010Berry} (here, $P$ is the inversion symmetry).
Two-dimensional \moire materials are ideal for realizing these conditions since constituent layers with specific symmetries can be combined in a controlled way to engineer both the electronic band structure and its topology.
Indeed, the AHE was realized in different $C_{2z}$-broken \moire platforms, including graphene-based systems \cite{Sharpe2019Emergent, Serlin2020Intrinsic, Chen2020Tunable, Polshyn2020electrical, Chen2021Electrically, Tschirhart2021Imaging, He2021competing, Polshyn2022Topological, Zhang2023local, Tseng2022anomalous, Kuiri2022, he2021chiralitydependent, Lin2022spinorbit, polski2022hierarchy, zhang2023valley, Chen2022tunable} and transition-metal dichalcogenides \cite{li2021quantum, anderson2023programming, Cai2023Signatures, zeng2023integer, foutty2023mapping, park2023observation}, where at least one of the van der Waals layers intrinsically breaks $C_{2z}$.
These systems rely on a single \moire to generate flat bands and strong correlations, while layers with broken $C_{2z}$ give rise to Berry curvature.

\begin{figure*}
    \centering
    \includegraphics[width=4.72in]{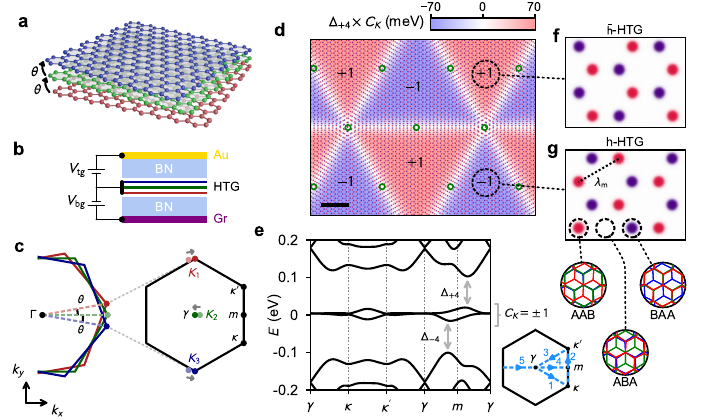}
    \caption{\figtitle{Helical trilayer graphene.}
    \panel{a}, Schematic of the magic-angle HTG structure comprising three layers of graphene rotated in the same direction by the same twist angle, $\theta\approx\SI{1.8}{\degree}$.
    \panel{b}, Circuit diagram of HTG surrounded by two hBN dielectric layers and top and bottom gate electrodes (Au and graphite, Gr) kept at electric potentials $V_\text{tg}$, $V_\text{bg}$ relative to HTG.
    \panel{c}, The rotated Brillouin zones of the three monolayers (left).
    Without lattice relaxation, the three Dirac points (light-colored dots) lie on an arc.
    Including lattice relaxation (grey arrows), in the h-HTG and \hb-HTG domains they lie on a straight line, forming a periodic \moire Brillouin zone.
    \panel{d}, \Moire structure of HTG including lattice relaxation.
    Crimson and purple dots represent AA stacking of the top and bottom pairs of proximate layers, respectively.
    The small atomic lattice relaxation is enough to form \moire-periodic domains (h-HTG and \hb-HTG) hosting topological flat bands with large gaps, $\Delta_{\pm 4}$, to remote bands. Scale bar is \SI{50}{nm}.
    Background color represents $\Delta_{+4} \times C_K$ calculated for the local shift $\delta$ between the two \moire lattices, where $C_K$ is the total Chern number per spin of the pair of flat bands in valley $K$.
    Green circles indicate centers of approximate $C_{2z}$ rotation symmetry.
    \panel{e}, Local non-interacting band structure in the h-HTG domain for valley $K$.
    \panel{f-g}, \Moire arrangement in the \hb-HTG (\panel{f}) and h-HTG (\panel{g}) periodic domains, where $C_{2z}$ is broken on the \moire scale.
    Circular callouts show the local atomic stacking of the three graphene layers in different regions.
    }
    \label{fig:setup}
\end{figure*}

Three layers of graphene, however, twisted with two independent angles, $\theta_{12}$ and $\theta_{23}$, between the top two layers (1 and 2) and bottom two layers (2 and 3), respectively, unlock a new engineering degree of freedom by combining more than one \moire lattice.
While much of the two-angle space ($\theta_{12},\theta_{23}$) is filled with mutually incommensurate \moire lattices forming \moire quasicrystals \cite{uri2023superconductivity}, some angle combinations will give rise to periodic domains on the \moire scale.
Some commensurate combinations of twist angles, $p\theta_{12}=q\theta_{23}$ ($p,q$ integers) \cite{mora2019flatbands,Foo2023extendedmagic,popov2023magicAngleButterfly}, may locally escape quasiperiodicity, particularly once lattice relaxation is taken into account.
Here, we demonstrate \moire engineering of local periodicity with $(p,q)=(1,1)$ equi-angle HTG, which we refer to as HTG throughout.
In HTG, large \moire-periodic domains are formed by lattice relaxation, supporting electronic Bloch bands, emergent broken symmetries, and Chern numbers defined within each domain.
Although HTG is made purely of $C_{2z}$-symmetric components and is globally $C_{2z}$-symmetric, within the periodic domains the system breaks $C_{2z}$ on the \moire scale, generating Berry curvature and non-trivial topology.
Furthermore, the \moire bands can be sufficiently flat to spontaneously break time-reversal symmetry (TRS), leading to our observation of the AHE at certain fillings.
Our observation of the AHE in HTG brings to light the importance of local symmetries on length scales comparable to the interparticle distance, $n^{-1/2}$.

\section*{Helical trilayer graphene}
The three graphene layers in HTG, rotated sequentially by the same angle $\theta$ (\figlbl~\ref{fig:setup}\panel{a}), form two \moire patterns between adjacent layer pairs.
The two \moire lattices share the same lattice constant, $\lambda_\text{m} \approx a/\theta \approx \SI{7.8}{nm}$ for $\theta = \SI{1.8}{\degree}$, and are misoriented by $\theta$ (here, $a=\SI{0.246}{nm}$ is the atomic lattice constant of graphene).
The misorientation between the two \moire patterns produces a position-dependent relative shift between them that is approximately periodic.
This \supermoire (or \moire of \moire) pattern has a lattice constant $\lambda_\text{sm} \approx \lambda_\text{m}/\theta \approx \SI{250}{nm}$.
In the absence of lattice relaxation, the two \moire patterns are mutually incommensurate and the effective structure for low-energy electrons is a \moire quasicrystal \cite{uri2023superconductivity}.
Lattice relaxation on the \moire scale is most prominent at small angles, $\theta \lesssim \SI{1}{\degree}$, and is typically less important for larger angles \cite{yoo2019atomic}.
However, even at angles as large as $\theta = \SI{1.8}{\degree}$, relaxation can have a profound effect on the \supermoire scale, favoring certain relative shifts between the two \moire patterns.
The relaxed structure of HTG forms large \moire-periodic domains, termed h-HTG and \hb-HTG, that are related by a $C_{2z}$ transformation \cite{devakul2023magicangle,nakatsuji2023multiscale}.
Figures~\ref{fig:setup}\panel{d,f,g} show the calculated relaxed \moire structure (Methods~\ref{ssec:relaxation}).
Importantly, within each periodic domain, $C_{2z}$ symmetry is spontaneously broken, allowing for Berry curvature and non-trivial topology.

\begin{figure*}
    \centering
    \includegraphics[width=7in]{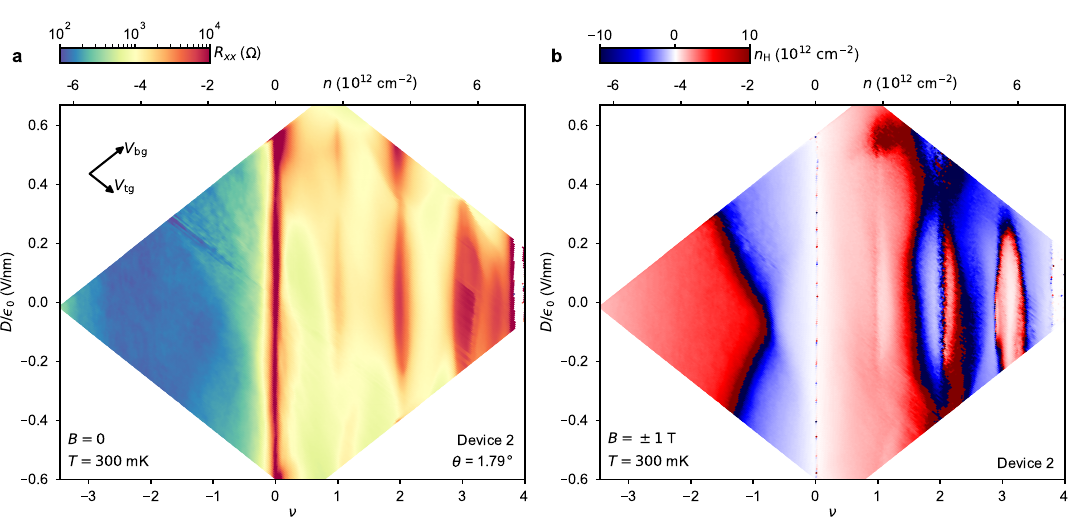}
    \caption{\figtitle{Strong electronic interactions.}
    \panel{a}, $R_{xx}$ versus $\nu$ and $D$ measured at $B=0$ and $T=\SI{300}{mK}$, showing resistance peaks at charge neutrality ($\nu=0$), at the \moire band gap ($\nu=4$), and at the correlated states ($\nu=1,2,3$ and $7/2$).
    \panel{b}, Hall density $n_\mathrm{H}$ versus $\nu$ and $D$, acquired from antisymmetrizing $R_{yx}$ at $B = \pm\SI{1}{T}$ and $T=\SI{300}{mK}$.
    }
    \label{fig:nD}
\end{figure*}

The local non-interacting band structure within a domain is described by a pair of narrow-bandwidth topological bands (per spin and valley), touching at three Dirac points ($\kappa$, $\kappa'$, and $\gamma$) and separated by large gaps of order $\SI{70}{meV}$ from the remote bands (Figs.~\ref{fig:setup}\panel{c,e}). 
Each pair of flat bands carries a total valley-contrasting Chern number $C_{K,K'}=\pm1$ (\figlbl~\ref{fig:setup}\panel{e}) \cite{devakul2023magicangle,nakatsuji2023multiscale}.
Electronic interactions are predicted to give rise to correlated insulators at integer fillings, in the form of spontaneous flavor ferromagnetism in spin ($\uparrow$,$\downarrow$), valley ($K$, $K'$), and Chern-sublattice ($A$, $B$) space with Chern numbers $C=\pm(1,-2)$~\cite{kwan2023strongcoupling}.
We will refer to the Chern-sublattice basis as sublattice for short hereinafter.
In states where h-HTG and \hb-HTG bands have different Chern numbers per spin or valley the network of domain walls between them hosts gapless edge modes \cite{devakul2023magicangle,nakatsuji2023multiscale}.
Relaxation into periodic \moire domains comes at a cost of increased \moire aperiodicity within the domain walls \cite{devakul2023magicangle}.
The overall structure of HTG is therefore a triangular tiling of \moire-periodic domains (on the scale of a few hundred nanometers) that are separated by gapless \moire-aperiodic domain walls (\figlbl~\ref{fig:setup}\panel{d}).
The global structure has approximate $C_{2z}$-symmetry, as h-HTG and \hb-HTG are $C_{2z}$ counterparts (see Figs.~\ref{fig:setup}\panel{f,g}).

In the absence of TRS breaking, both valleys are equally occupied and the total Chern number is zero.
Electron-electron interactions facilitated by the quenched kinetic energy of the flat bands can lift the flavor degeneracy, similar to magic-angle graphene \cite{Cao2018Unconventional,Zondiner2020cascade, Wong2020cascade}, spontaneously breaking TRS and yielding a net Chern number within each periodic domain \cite{Sharpe2019Emergent,Serlin2020Intrinsic}.

\section*{Correlated states}
To probe the transport properties of HTG, we constructed dual-gated devices encapsulated by hexagonal boron nitride (hBN) (Methods~\ref{ssec:fab}, \extlbl~\ref{fig:devices}).
By applying voltages to the top and bottom gates ($V_\text{tg}$ and $V_\text{bg}$, respectively, see \figlbl~\ref{fig:setup}\panel{b}), we controlled the electron density, $n$, and the perpendicular electric displacement field, $D$, independently, while performing four-terminal magneto-transport measurements (Methods~\ref{ssec:transport}).
The data shown throughout were measured using Device~2 unless stated otherwise.

\begin{figure*}
    \centering
    \includegraphics[width=7in]{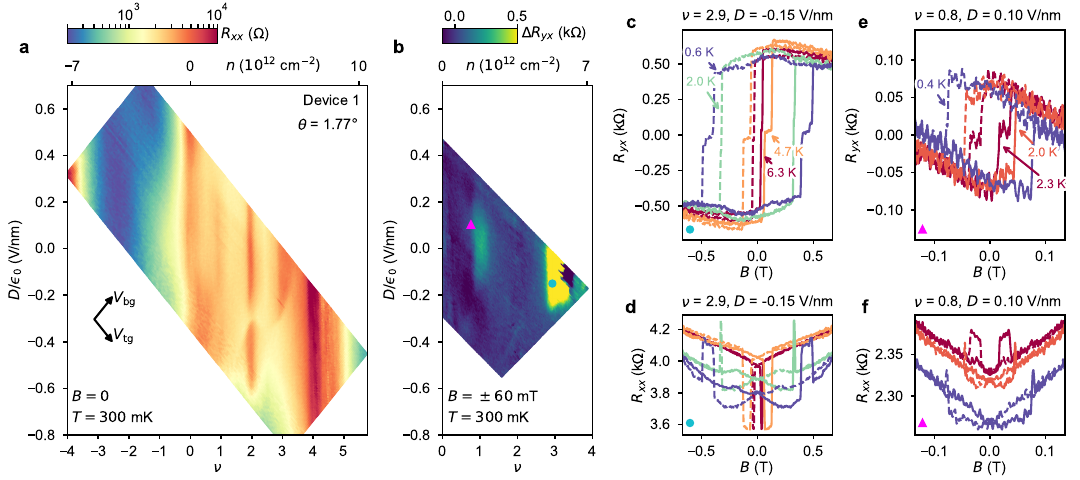}
    \caption{\figtitle{Device 1 characterization.}
    \panel{a}, $R_{xx}$ versus $n$ and $D$, showing resistance peaks at charge neutrality ($\nu=0$), at the \moire band gaps ($\nu=\pm4$), and at the correlated states at $\nu=1,2,3$.
    \panel{b}, Field-trained $\Delta R_{yx}$ measured at $T=\SI{300}{mK}$ and $B=\SI{\pm 60}{mT}$ versus $\nu$ and $D$.
    Hot spots near $\nu=1,3$ indicate AHE.
    \panel{c,d}, Field-antisymmetrized $R_{yx}$ and field-symmetrized $R_{xx}$ taken at $\nu=2.9$ (cyan circle in \panel{b}) and $D/\epsilon_0=\SI{-0.15}{V/nm}$ while sweeping $B$ up (solid) and down (dashed) at different temperatures as indicated.
    Temperature colorcode in \panel{d} is identical to \panel{c}.
    \panel{e,f}, Same as \panel{c,d}, taken at $\nu=0.8$ and $D/\epsilon_0=\SI{0.1}{V/nm}$ (pink triangle in \panel{b}).}
    \label{fig:device1}
\end{figure*}

\begin{figure*}
    \centering
    \includegraphics[width=7in]{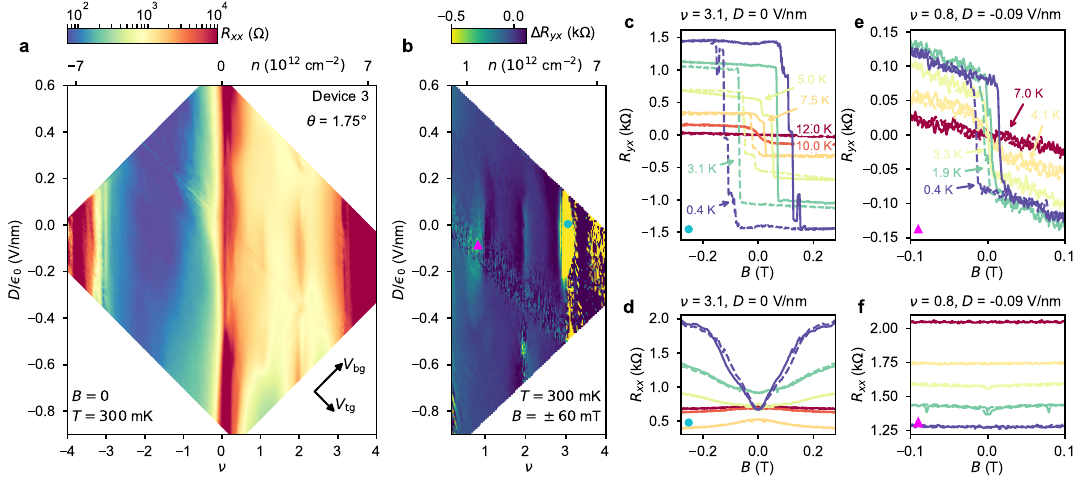}
    \caption{\figtitle{Device 3 characterization.}
    \panel{a}, $R_{xx}$ versus $n$ and $D$, showing resistance peaks at charge neutrality ($\nu=0$), at the \moire band gaps ($\nu=\pm4$), and at the correlated states at $\nu=1,2,3$.
    The contact resistance becomes very large when $\nu\gtrsim 3.2$, leading to artifacts in the data.
    \panel{b}, Field-trained $\Delta R_{yx}$ measured at $T=\SI{300}{mK}$ and $B=\SI{\pm 60}{mT}$ versus $\nu$ and $D$.
    Hot spots near $\nu=1,3$ indicate AHE.
    \panel{c,d}, Field-antisymmetrized $R_{yx}$ and field-symmetrized $R_{xx}$ taken at $\nu=3.1$ (cyan circle in \panel{b}) and $D/\epsilon_0=0$ while sweeping $B$ up (solid) and down (dashed) at different temperatures as indicated.
    The temperature color code in \panel{d} is identical to \panel{c}.
    \panel{e,f}, Same as \panel{c,d}, taken at $\nu=0.8$ and $D/\epsilon_0=\SI{-0.09}{V/nm}$ (pink triangle in \panel{b}).}
    \label{fig:device3}
\end{figure*}

Figure~\ref{fig:nD}\panel{a} shows the longitudinal resistance $R_{xx}$ versus $n$ and $D$ measured at $B=0$ and $T=\SI{300}{mK}$, where $B$ is the applied out-of-plane magnetic field.
The large resistance peak at $n=\SI{7.45e12}{cm^{-2}}$ indicates the gap between the flat and remote \moire bands at a filling of four electrons per \moire unit cell, $\nu=n A_\mathrm{uc}=4$, where $A_\mathrm{uc}$ is the area of the \moire unit cell.
This reflects a twist angle of $\theta=\SI{1.79}{\degree}$ (Methods~\ref{ssec:twist}).
Additional resistance peaks appear at integer fillings $\nu=1,2,$ and $3$, indicative of flavor-symmetry-broken correlated electronic states.
We observed similar behavior in two more devices -- see \extlbls~\ref{fig:device1}\panel{a} and \ref{fig:device3}\panel{a} (for the appearance of correlated states in HTG devices with twist angles away from the magic angle see Methods~\ref{ssec:twistdependence} and Table~\ref{table:devices}).
\figlbl~\ref{fig:nD}\panel{b} shows the Hall density, $n_\mathrm{H}$ (Methods \ref{ssec:Halldensity}), measured versus $n$ and $D$.
For electron-doping, $n>0$, and small $D$, $n_\mathrm{H}$ shows ``reseting'' behavior around $\nu=1$ and $2$, where $n_\mathrm{H}$ drops towards zero \cite{Zondiner2020cascade, Wong2020cascade}.
Additionally, we find Van Hove singularities (VHSs) near $\nu=1.5$ and 2.2 \cite{Park2021Tunable}.
The rightmost oval-shaped discontinuity in $n_\mathrm{H}$ near $\nu=3$ does not reflect VHSs.
Rather, they are the result of the strong AHE in this region (see details below) that we do not account for in our extraction of $n_\mathrm{H}$.

Both the $R_{xx}$ and $n_\mathrm{H}$ maps reveal pronounced electron-hole asymmetry.
Correlated states only appear on the electron side, an observation we find consistently in all of our HTG devices (see \extlbls~\ref{fig:device1}, \ref{fig:device3}, \ref{fig:otherangles}, and \exttbllbl~\ref{table:devices}).
Our current microscopic theory does not account for this strong asymmetry.
We can, however, use the shape of the valence band VHS in the measured Hall density map to infer effective microscopic parameters for our model.
Compared to the electron side, the hole side is better described (phenomenologically) by a larger ratio of renormalized Fermi velocity to interlayer hopping strength (see Methods \ref{ssec:microscopic}), resulting in a larger bandwidth and diminished correlation effects at $\theta\approx\SI{1.8}{\degree}$.

\section*{Anomalous Hall effect}

\begin{figure*}
    \centering
    \includegraphics[width=7in]{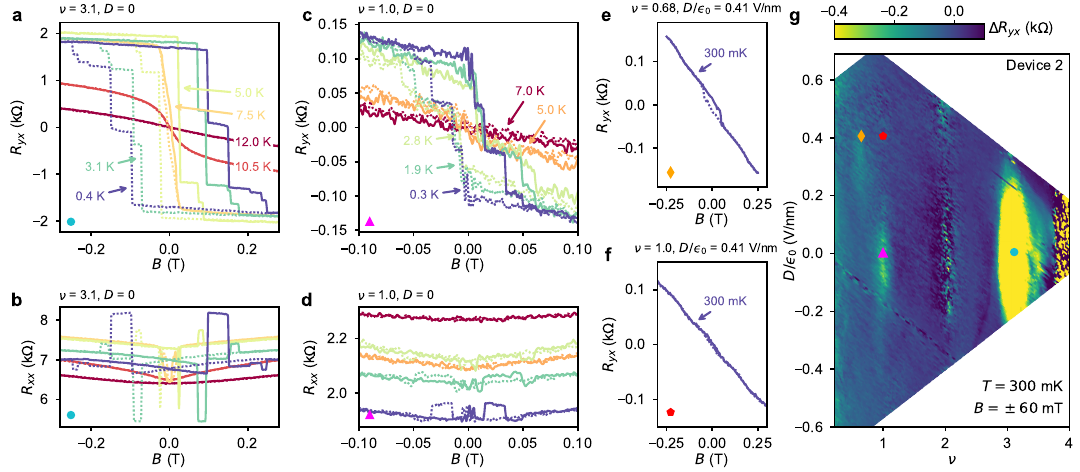}
    \caption{\figtitle{Anomalous Hall effect.}
    \panel{a,b}, Field-antisymmetrized $R_{yx}$ and field-symmetrized $R_{xx}$ taken at $\nu=3.1$ and $D=0$ (cyan circle in \panel{g}) while sweeping $B$ up (solid) and down (dashed) at different temperatures as indicated.
    Temperature color code in \panel{b} is identical to \panel{a}.
    \panel{c,d}, Same as \panel{a,b}, taken at $\nu=1$ (pink triangle in \panel{g}).
    \panel{e}, Same as \panel{a}, taken at $\nu=0.62$, $D/ \epsilon_0 = \SI{0.41}{V/nm}$ (orange diamond in \panel{g}), and $T=\SI{300}{mK}$ showing AHE near $\nu=2/3$.
    \panel{f}, Same as \panel{e}, taken at the same $D$ and $T$ but at $\nu=1$ (red pentagon in \panel{g}) showing no AHE.
    \panel{g}, Field-trained $\Delta R_{yx}$ measured at $T=\SI{300}{mK}$ and $B=\SI{\pm 60}{mT}$ after training at $B=\pm\SI{1}{T}$ versus $\nu$ and $D$.
    Hot spots near $\nu=1,3$ and $\nu \approx 2/3$ indicate AHE.
    }
    \label{fig:AHE}
\end{figure*}

Figures~\ref{fig:AHE}\panel{a-f} show the measured field-symmetrized $R_{xx}$ and field-antisymmetrized $R_{yx}$ as we sweep $B$ up and down at different temperatures (see Methods \ref{ssec:symmetrization} for a description of the symmetrization and antisymmetrization procedure).
We observe nonzero Hall resistance at $B=0$ accompanied by pronounced magnetic hysteresis consistent with ferromagnetism in the vicinity of both $\nu=1$ and 3.
Considering the small intrinsic spin-orbit coupling in graphene, the ferromagnetism in our system is almost certainly of orbital origin rather than spin alone \cite{Huang2021current, sharpe2021evidence}, as was shown \cite{Tschirhart2021Imaging, Grover2022} in other graphene-based \moire systems in which AHE was previously reported.
Combined with the topological flat bands (\figlbl~\ref{fig:setup}\panel{e}) \cite{devakul2023magicangle,nakatsuji2023multiscale,guerci2023chern}, this points to orbital ferromagnetism driven by Berry curvature in valley-polarized states \cite{kwan2023strongcoupling}.
While $R_{yx}$ is not quantized in our measurements, we would only expect to observe quantization from a single Chern domain.
The magnitude of $|R_{yx}| \gtrsim \SI{1.5}{k\Omega}$ at $B=0$ in our experiment reflects the net effect of multiple domains and the network of gapless domain walls between the $R_{yx}$ contacts.
We find multiple Barkhausen jumps and hysteresis in $R_{yx}$-versus-$B$ loops that persists up to a temperature of $T_\mathrm{hys}\approx \SI{7.5}{K}$. 
The $R_{yx}$ discontinuity at $B=0$ disappears at a Curie temperature of $T_\text{C} \approx \SI{10.5}{K}$ (see also Methods \ref{ssec:Curie} and \extlbl~\ref{fig:Arrott}), which is the highest reported among graphene-based \moire systems (see \exttbllbl~\ref{table:Tc}).

Notably, we find the AHE at $\nu=1,3$ in two additional devices with similar twist angles (see \extlbl~\ref{fig:device1} and \extlbl~\ref{fig:device3}), demonstrating the robustness of the AHE in HTG.
Apart from non-trivial band topology, another key requirement for interaction-driven orbital magnetism is that the ground state favored by strong correlations spontaneously breaks TRS and has a net Chern number.
In twisted monolayer-bilayer graphene (tMBG), previous studies suggested that a close competition exists between different many-body ground states, including a valley-polarized state that breaks TRS and an intervalley-coherent state that preserves it \cite{Polshyn2020electrical, He2021competing}.
This potentially makes the AHE in tMBG more sensitive to strain and twist angle disorder \cite{Zhang2023local, Lau2022Reproducibility}.
In contrast, the robustness of the AHE in HTG suggests that valley-polarized states are strongly favored at odd fillings.
This is in alignment with strong-coupling theory of HTG \cite{kwan2023strongcoupling}.

\begin{figure*}
    \centering
    \includegraphics[width=7in]{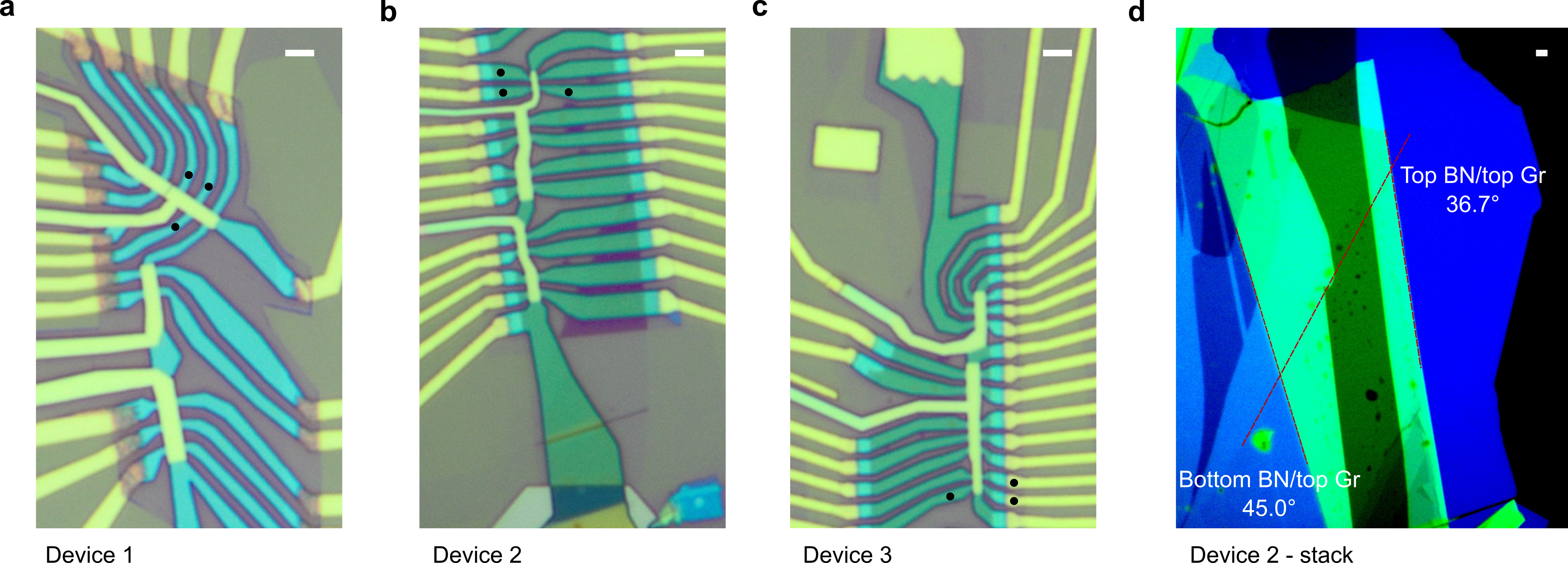}
    \caption{\figtitle{Optical micrographs of HTG devices.}
    \panel{a}, Device 1 -- a secondary device with $\theta=\SI{1.77}{\degree}$.
    \panel{b}, Device 2 -- our main device with $\theta=\SI{1.79}{\degree}$.
    \panel{c}, Device 3. This device shares the van der Waals heterostructure with Device 2.
    $R_{xx}$ and $R_{yx}$ contacts are indicated by black dots for all devices.
    \panel{d}, Contrast-enhanced optical micrograph of Device 2 after stacking. The crystallographic edges of the top hBN, bottom hBN, and top monolayer graphene are highlighted, showing no accidental alignment between hBN and HTG.
    All scale bars are $\SI{2}{\mu m}$.
    \label{fig:devices}
    }
\end{figure*}

Figure~\ref{fig:AHE}\panel{g} provides an overview of the AHE in our system by plotting the difference $\Delta R_{yx}$ between $R_{yx}$ taken at $B=\pm\SI{60}{mT}$ after training at high fields, $\pm \SI{1}{T}$, respectively.
In the ranges $0.9 \lesssim \nu \lesssim 1.1$ and $2.8 \lesssim \nu \lesssim 3.3$, we find nonzero $\Delta R_{yx}$ indicating AHE, corroborated by the $B$-sweep hysteresis loops.
In contrast, the correlated state near $\nu =2$ shows no AHE, indicating a state that preserves TRS.
At this filling fraction theory points to a quantum valley-Hall state within the periodic domains \cite{kwan2023strongcoupling}.
The approximate symmetry of the AHE about $D=0$ indicates that it does not rely on aligning the graphene trilayer to a substrate (see also \extlbl~\ref{fig:devices}\panel{d}).

There is a weaker $\Delta R_{yx}$ hot spot near $\nu=2/3$ and $D/\epsilon_0=\SI{0.3}{V/nm}$ (\figlbl~\ref{fig:AHE}\panel{g}).
Figure~\ref{fig:AHE}\panel{e} shows $R_{yx}$ versus $B$ measured at $\nu=0.62$, $D=\SI{0.41}{V/nm}$ showing AHE.
We note that at the same $D$-field there is no AHE at $\nu=1$ (\figlbl~\ref{fig:AHE}\panel{f}), hence the AHE at fractional filling is a distinct state.
This observation may indicate the presence of a topological charge density wave or fractional Chern insulator (see Methods~\ref{ssec:HF}).
The latter was predicted in HTG at D=0 \cite{devakul2023magicangle};
However, it is unclear whether the HTG bands favor fractional Chern insulators at large D fields.
Thus, further investigation is required to identify the ground state.
The appearance of the $\nu=2/3$ feature only at $D>0$ may be due to a slight difference in the effective screening from the top and bottom gates.

\begin{figure}
    \centering
    \includegraphics[width=3.27in]{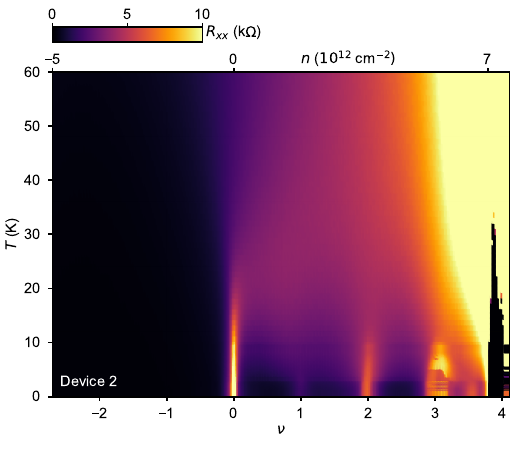}
    \caption{\figtitle{Temperature dependence.}
    $R_{xx}$ (raw data, not field-symmetrized) versus $\nu$ and $T$ of Device~2 at $D=0$ and $B=0$.
    The jumps in resistance near $\nu=3$ reflect the AHE of different magnetic states combined with $R_{yx}$ mixing.
    A pronounced electron-hole asymmetry is demonstrated.}
    \label{fig:gaps}
\end{figure}

\begin{figure*}
    \centering
    \includegraphics[width=5.2in]{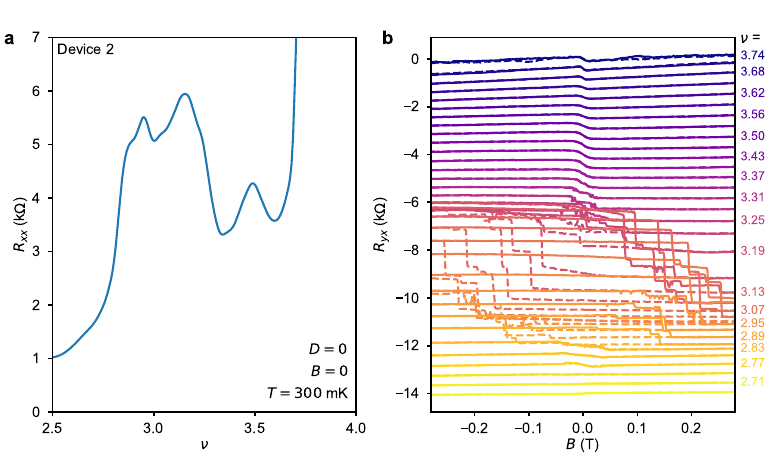}
    \caption{\figtitle{Correlated state at $\nu=7/2$.}
    \panel{a}, $R_{xx}$ versus $\nu$, measured on Device~2 at $D=0$, $B=0$, and $T=\SI{300}{mK}$. At $\nu=7/2$ we find a resistance peak distinct from the one at $\nu=3$.
    \panel{b}, Waterfall plot of antisymmetrized $R_{yx}$ taken by sweeping $B$ up (solid) and down (dashed) as the fast axis at $D=0$ and different $\nu$, as indicated on the right of every other curve.
    The AHE persists beyond $\nu=7/2$.
    }
    \label{fig:waterfall_nu3}
\end{figure*}

Lastly, there is an isolated resistive state centered on $\nu=7/2$ and $D=0$ (see \figlbl~\ref{fig:nD}\panel{a}), that emerges as a distinct feature at low temperatures (\extlbl~\ref{fig:gaps}), indicating a symmetry-broken phase.
Furthermore, we find AHE that extends from $\nu=3$ to beyond $\nu=7/2$ (\extlbl~\ref{fig:waterfall_nu3}\panel{b}) indicating the nontrivial topology of this state.
At this filling, our Hartree-Fock calculations find closely competing topological states that include charge density waves \cite{Dong2023manybody} and a tetrahedral antiferromagnet (Methods~\ref{ssec:HF}).

\section*{Indications of topological phase transitions}
In \figlbl~\ref{fig:nD}\panel{a} the resistive peak at $\nu=1$ centered at $D=0$ disappears at $\abs{D/\epsilon_0}\approx \SI{0.35}{V/nm}$ and reappears at higher $|D|$, suggestive of a phase transition involving a gap closure and re-opening.
We do not find evidence for AHE at $\abs{D/\epsilon_0}>\SI{0.35}{V/nm}$, hence the high-$\abs{D}$ phases preserve TRS.
A leading theoretical possibility suggested by recent Hartree-Fock calculations of HTG \cite{kwan2023strongcoupling} is that at a critical displacement field $D_\mathrm{c}$ band inversion leads to a topological phase transition, with trivial bands emerging at $|D|>D_\mathrm{c}$.
While a gapped intervalley-coherent state is also expected to show similar transport signatures, this possibility is unlikely \cite{devakul2023magicangle,kwan2023strongcoupling}.
We observe similar behavior of $R_{xx}$ near $\nu=2$, although at this filling both the high- and low-$\abs{D}$ phases preserve TRS, also consistent with theory \cite{kwan2023strongcoupling}.

\section*{Electrical switching of Chern domains}
\begin{figure}
    \centering
    \includegraphics[width=3.4in]{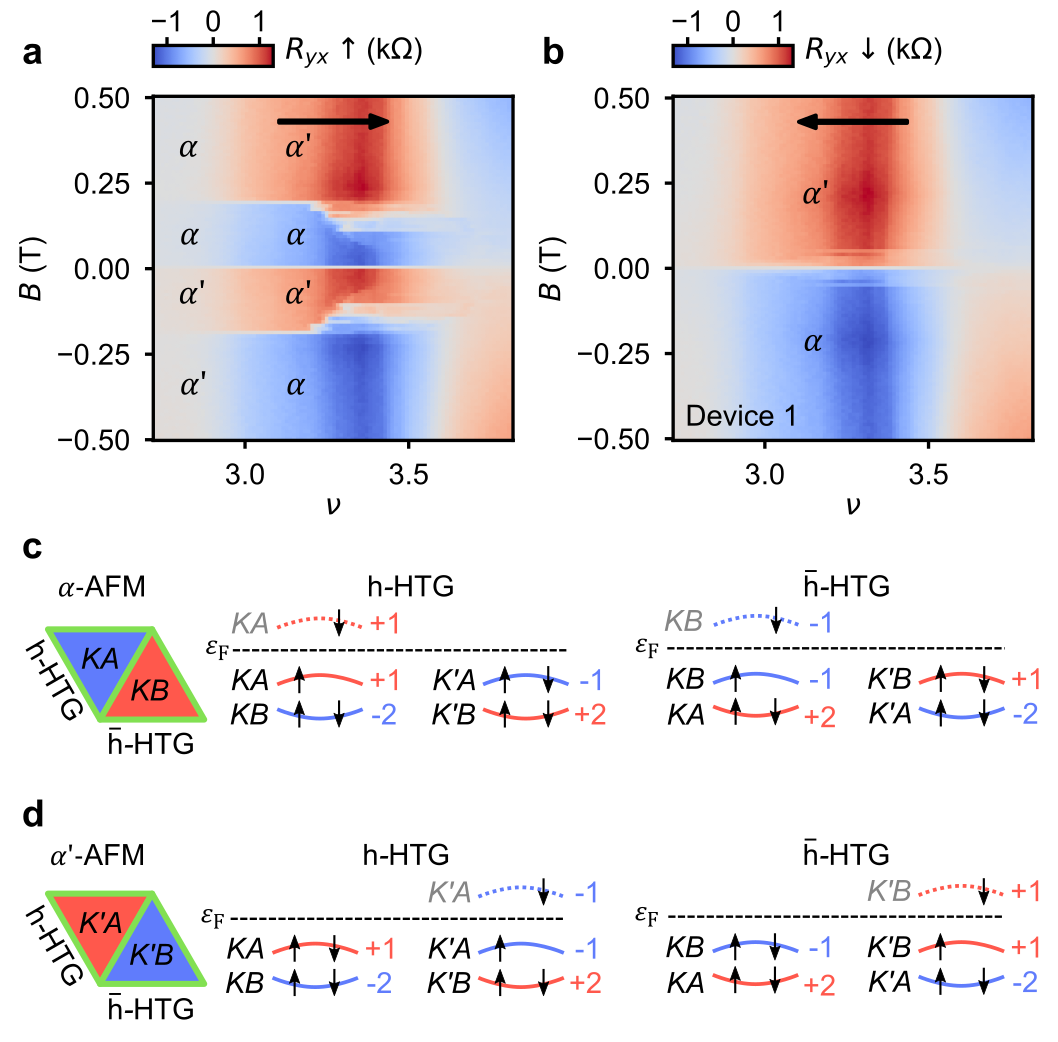}
    \caption{\figtitle{Density-induced switching and Chern mosaic at $\nu=3$.}
    \panel{a}, Field-antisymmetrized $R_{yx}$ versus $\nu$ and $B$ while sweeping $\nu$ up as fast axis, measured on Device 1 at $D/\epsilon_0=\SI{-0.15}{V/nm}$.
    The theoretical $\alpha$ and $\alpha'$ AFM configurations are indicated.
    \panel{b}, Same as \panel{a} only sweeping $\nu$ down.
    \panel{c,d}, Schematic of the flavor-energetics in h-HTG and \hb-HTG domains in the interacting picture for $\nu=3$, describing the two AFM orders, $\alpha$ (\panel{c}) and $\alpha'$ (\panel{d}).
    At integer $\nu$, interactions gap the Dirac points, forming eight bands with spin, valley, and sublattice flavors.
    The Chern numbers per band are indicated.
    At $\nu=3$ seven flavors are filled, and one is empty (dashed).
    The two triangles represent the h-HTG and \hb-HTG domains (empty flavor is indicated).
    Red (blue) indicates total Chern number 1 (-1) while green indicates gapless regions at $\nu=3$.
    \label{fig:mosaic}
    }
\end{figure}

By sweeping density in a fixed small magnetic field, $B\lesssim\SI{0.2}{T}$, we find that the sign of the AHE for a given $\nu$ depends on the sweep direction (Figs.~\ref{fig:mosaic}\panel{a,b}).
This type of switching was suggested to be a result of a competition between the different contributions to the orbital Zeeman energy, $-\mathbf{M} \cdot \mathbf{B}$ \cite{Zhu2020voltage}, where $\mathbf{M}$ is the total orbital magnetization, directed out of the plane.
$M$ has two contributions \cite{Xiao2010Berry}: the self-rotation magnetization, $M_\text{SR}$, due to the self-rotation of the electronic wavepacket, and the Chern magnetization, $M_\text{C}$, due to the center-of-mass motion of the wavepacket.
The two can have opposite signs and they depend strongly on the chemical potential, allowing the total magnetization per valley to change sign as the density is swept.
A calculation of the orbital magnetization $M(\nu)$ for the h-HTG domain, based on Hartree-Fock bands calculated at $\nu =3$, is presented in Methods~\ref{ssec:magnetizationreversal}, demonstrating the sign change required for the above switching mechanism.

The hysteresis in $R_{yx}(\nu)$ shown in Figs.~\ref{fig:mosaic}\panel{a,b} exists over a large density range.
Notably, it is present only at low magnetic fields and disappears abruptly as $\abs{B}$ is increased above a critical value $B_\mathrm{c}\sim\SI{0.2}{T}$.
Moreover, when sweeping $\nu$ up (Fig.~\ref{fig:mosaic}\panel{a}), $R_{yx}$ in the high positive field regime ($B > B_\mathrm{c}$) is almost identical to $R_{yx}$ in low negative field ($0<B< B_\mathrm{c}$).
This phenomenology is strikingly similar to what was previously observed \cite{Grover2022} in near-commensurate hBN-aligned magic-angle TBG and is distinct from the phenomenology observed in other orbital Chern insulators \cite{Polshyn2020electrical, Chen2021Electrically, Tseng2022anomalous}.
In the former, local magnetic imaging revealed a spatial pattern of domains with different Chern numbers in the low-field regime \cite{Grover2022}.
Our observation of a similar phenomenology in $R_{yx}(\nu,B)$ is not surprising, given the coexistence of h-HTG and \hb-HTG domains that may have different Chern numbers at integer filling factors \cite{kwan2023strongcoupling}.
Below, we give a possible explanation for the observed unique hysteresis pattern.
We note, however, that the multiple domains, gapless domain walls, and the large number of degrees of freedom may give rise to complex magnetotransport behavior.
Hence, the picture described below may be incomplete.

At $\nu=3$, as a result of interaction-induced flavor polarization, seven of the eight flavors are filled, and one is empty.
Figure ~\ref{fig:mosaic}\panel{c} illustrates such a scenario.
Because both $A$/$B$ sublattices and h-HTG/\hb-HTG domains are related by $C_{2z}$, we expect that if the unfilled band in h-HTG is, say $B$-sublattice polarized, then in \hb-HTG it would be $A$-sublattice polarized.
Therefore, the sublattice polarization of the empty flavor is domain contrasting (Figs.~\ref{fig:mosaic}\panel{c,d}).
We are now left with the assignment of valley and spin to the empty flavor in each domain.
Because of the vanishingly small spin-orbit coupling in graphene we ignore the spin in the following discussion.
At a nonzero magnetic field, one naively expects the system to minimize orbital-Zeeman energy by aligning the orbital magnetization of all the domains with the applied field.
This can be achieved by polarizing h-HTG and \hb-HTG domains to opposite valleys.
However, due to the high valley-domain-wall energy cost \cite{kwan2023strongcoupling}, this arrangement is disfavored.
Instead, all the domains polarize collectively to the same valley, breaking TRS, and forming an inter-domain antiferromagnetic (AFM) order (Figs.~\ref{fig:mosaic}\panel{d,e}) -- a mosaic of Chern domains \cite{Grover2022}.
We denote the two possible AFM configurations as $\alpha$ and $\alpha'$, corresponding to $K$ and $K'$ valley polarizations, respectively (Figs.~\ref{fig:mosaic}\panel{c,d}).
The inevitable imbalance between h-HTG and \hb-HTG domains in a real sample results in a small net orbital magnetization $M_\mathrm{net}$ that couples to the applied magnetic field via orbital-Zeeman energy, $\delta \varepsilon = -\mathbf{M}_\mathrm{net}\cdot \mathbf{B}$.
The uncompensated AFM can therefore be switched between $\alpha$ and $\alpha'$ if $B$ is applied anti-parallel to $M_\mathrm{net}$ and the Zeeman energy exceeds some coercive energy threshold, $|\delta \varepsilon| > \Delta_\mathrm{c}$.

We now describe Fig.~\ref{fig:mosaic}\panel{a}.
At $\nu<3$ and $B>0$, $M_\mathrm{net}$ is aligned with $B$ and system is in the $\alpha$ configuration.
As the density is swept up through the $\nu=3$ correlated gap, the total magnetization per domain switches sign abruptly (see Methods~\ref{ssec:magnetizationreversal}) and so does $M_\mathrm{net}$.
At a high magnetic field, $B>B_\mathrm{c}$, $|\delta \varepsilon|$ exceeds the coercive energy 
$\Delta_\mathrm{c}$, and the system collectively switches to the $\alpha'$ configuration.
In contrast, at low fields, $0 \leq B < B_\mathrm{c}$, as $\nu$ is swept through the gap, $|\delta \varepsilon| < \Delta_\mathrm{c}$ and the AFM order cannot be flipped although $M_\mathrm{net}$ is anti-parallel to the field.
This explains the opposite $R_{yx}$ at high and low fields in Fig.~\ref{fig:mosaic}\panel{a}.
Using the same argument, we can also understand why the small negative field regime, $-B_\mathrm{c} < B < 0$, shows $R_{yx}$ similar to the positive high field regime, $B > B_\mathrm{c}$.
At small negative field, the system is in the $\alpha'$ configuration to align $M_\mathrm{net}$ with the field.
The system remains in the $\alpha'$ configuration also for $\nu>3$ because $|\delta \varepsilon| < \Delta_\mathrm{c}$.

We now describe Fig.~\ref{fig:mosaic}\panel{b} showing $R_{yx}$ while sweeping $\nu$ down.
Above some critical filling fraction, $\nu_\mathrm{c}$, the intradomain ferromagnetic order melts.
Sweeping down from $\nu>\nu_\mathrm{c}$ is akin to field-cooling a ferromagnet;
As the system crosses $\nu_\mathrm{c}$ and becomes ferromagnetic it chooses the energetically favorable AFM order which is $\alpha$ ($\alpha'$) for $B>0$ ($B<0$), see Fig.~\ref{fig:mosaic}\panel{b}.

Although the phenomenology we observe in Figs.~\ref{fig:mosaic}\panel{a,b} is similar to near-commensurate hBN-aligned magic-angle TBG \cite{Grover2022}, important differences exist between the two systems.
Further work is therefore required to determine the precise dynamics underlying the observed transport behavior in Figs.~\ref{fig:mosaic}\panel{a,b}.

\section*{Scale-dependent symmetries}
It is instructive to consider the symmetries of a locally-periodic system at the length scale of the inter-particle distance, $n^{-1/2}$.
For example, TBG is $C_{2z}$-symmetric at densities of order one electron per \moire unit cell, as embodied in the continuum approximation \cite{Bistritzer2011moire}.
This accounts for the scarcity of AHE observations in hBN-misaligned TBG.
In contrast, in hBN-aligned TBG, $C_{2z}$ is broken by the inequivalence between the alignment of boron and nitrogen atoms with the graphene \moire unit cell.
Since there exists no center of $C_{2z}$ rotation anywhere in the lattice, this symmetry is broken everywhere, promoting non-trivial topology in the \moire bands.
A similar situation exists in tMBG and in twisted transition-metal dichalcogenides.
HTG is qualitatively different.
$C_{2z}$ is broken on the \moire scale despite the global $C_{2z}$ symmetry of HTG.
The global symmetry is expressed by the existence of h-HTG and \hb-HTG domains with $C_{2z}$-related Chern numbers.

\section*{Outlook}
Our results demonstrate that engineering a \supermoire system to break $C_{2z}$ on the \moire scale can induce topological bands, despite the overall approximate $C_{2z}$ symmetry of the system.
We show that magic-angle HTG hosts flat topological bands that favor TRS-broken ground states at odd fillings.
The Chern domains and network of gapless edge modes, together with the high $T_\mathrm{C}$ and the high yield of samples exhibiting the AHE, establish HTG as an ideal platform for exploring orbital magnetism with Chern domain walls.
Owing to its relatively homogeneous charge distribution, near-ideal quantum geometry \cite{devakul2023magicangle}, and small interaction-induced dispersion \cite{kwan2023strongcoupling}, HTG is a promising platform for realizing exotic electronic phases \cite{devakul2023magicangle,Dong2023manybody}.
Our observation of an AHE at fractional filling factors, combined with the favorable conditions for zero-field fractional Chern insulating states \cite{devakul2023magicangle}, motivates further experimental and theoretical investigations of magic-angle HTG.

\clearpage
\onecolumngrid
\section*{Methods}

\subsection{Device fabrication}\label{ssec:fab}
The van der Waals heterostructures were assembled in two parts using the standard dry-transfer technique. First, a hBN flake and a few-layer graphene strip were picked up by a poly(bisphenol A carbonate) stamp. This bottom stack was released onto a $\SI{285}{nm}$ SiO$_2$/Si substrate, followed by 12-hour vacuum annealing at $\SI{350}{\degree C}$ to remove polymer residues. Then, tip cleaning was performed using the Contact Mode of a Bruker Icon XR atomic force microscope to further clean the surface. A monolayer graphene flake was cut into three pieces using a confocal laser-cut setup. A second poly(bisphenol A carbonate) stamp was used to pick a hBN flake and the three graphene pieces subsequently. Before picking up the second and third pieces of graphene, the stage was rotated by $\SI{1.8}{\degree}$ in the same direction to realize a helical stacking order. The pickup of graphene was done at room temperature to avoid the relaxation of the twist angle. The top stack was released onto the bottom stack at $\SI{150}{\degree C}-\SI{170}{\degree C}$.

The Hall bar was defined in a bubble-free region, identified under an atomic force microscope. Patterns were defined using an Elionix ELS-HS50 electron-beam lithography system. A metallic top gate (\SI{25}{nm} - \SI{65}{nm} Au with a \SI{2}{nm} - \SI{5}{nm} Cr or Ti adhesion layer) was deposited using a Sharon thermal evaporator. The device was connected using one-dimensional contacts (\SI{63}{nm} - \SI{75}{nm} Au with a \SI{2}{nm} - \SI{5}{nm} Cr adhesion layer) \cite{doi:10.1126/science.1244358}. Finally, the device was etched into a Hall bar geometry using reactive-ion etching.

\subsection{Electrical transport measurements}\label{ssec:transport}
Low-temperature electrical transport measurements were carried out in a helium-3 refrigerator with an $\SI{8}{T}$ perpendicular superconducting magnet and a base temperature of about $\SI{290}{mK}$.
A homemade twisted-pair copper tape filter with $\sim\SI{20}{MHz}$ cutoff frequency \cite{spietz2006twisted} was thermally anchored at base temperature to guarantee the electron temperature of the device is the same as the phonon temperature.
DC voltages were applied to the top and bottom gates using Keithley 2400/2450 source-measure units.
The AC excitation of $\SI{1}{nA}-\SI{10}{nA}$ at $\SI{16}{Hz}-\SI{24}{Hz}$ was applied using SR830 or SR860 lock-in amplifiers.
The corresponding AC currents and voltages were measured using SR830 or SR860 lock-in amplifiers, preamplified using DL-1211 current preamplifiers, and DL-1201 voltage preamplifiers.
The temperature was measured using a calibrated CX-1010-CU-HT-0.1L thermometer.
$n=(\epsilon_\mathrm{BN}\epsilon_0/e$)($V_\mathrm{bg}$/$d_\mathrm{bg} + V_\mathrm{tg}$/$d_\mathrm{tg}$) and $D=(\epsilon_\mathrm{BN}\epsilon_0/2$)($V_\mathrm{bg}$/$d_\mathrm{bg} - V_\mathrm{tg}$/$d_\mathrm{tg}$) define $n$ and $D$ relations to $V_\mathrm{bg}$ and $V_\mathrm{tg}$, where $\epsilon_\mathrm{BN}$ = 3 is the relative dielectric constant of hBN, $\epsilon_0$ is the vacuum permittivity, $e$ is the elementary charge, and $d_\mathrm{bg}$ ($d_\mathrm{tg}$) is the thickness of the bottom (top) hBN.

Dilution refrigerator measurements were performed in a Leiden Cryogenics CF-900 using a custom probe. 
The measurement lines are equipped with electronic filtering at the mixing chamber stage to obtain a low electron temperature in the device and reduce high-frequency noise. 
There are two stages of filtering: the wires are first passed through a cured mixture of epoxy and bronze powder to filter GHz frequencies, then low-pass RC filters mounted on sapphire plates filter MHz frequencies. 
Samples were mounted using a Kyocera custom 32-contact ceramic leadless chip carrier (drawing PB-44567-Mod with no nickel sticking layer under gold, to reduce magnetic effects).
Stanford Research Systems SR830 lock-in amplifiers with NF Corporation LI-75A voltage preamplifiers were used to perform four-terminal resistance measurements.
A \SI{1}{\giga\ohm} bias resistor was used to apply an AC bias current of up to \SI{5}{nA} RMS at a frequency of \SI{6.451}{Hz}.
Keithley 2400 source-measure units were used to apply voltages to the gates.

\subsection{Twist angle determination}\label{ssec:twist}
Band structure calculations for the h-HTG and \hb-HTG domains show large \moire band gaps at $\nu=\pm4$, while the domain walls remain gapless throughout the spectrum.
At $\nu=\pm4$, we expect the domain walls to form a metallic network shunting the gapped periodic domains and lowering the resistance at these fillings somewhat, compared with a homogeneous insulating system.
Nevertheless, we can still clearly identify resistive peaks at $\nu=4$ and Landau levels emerging from the band extrema.
We therefore use the features at $\nu=\pm4$ to extract a twist angle for each device, with $n_{\nu=\pm4} = \pm 8\sin^{2}\theta/\sqrt{3}a^2 \approx \pm 8\theta^2/\sqrt{3}a^2$, using $a=\SI{0.246}{nm}$ as the lattice constant for graphene.

\begin{figure}
    \centering
    \includegraphics[width=3.27in]{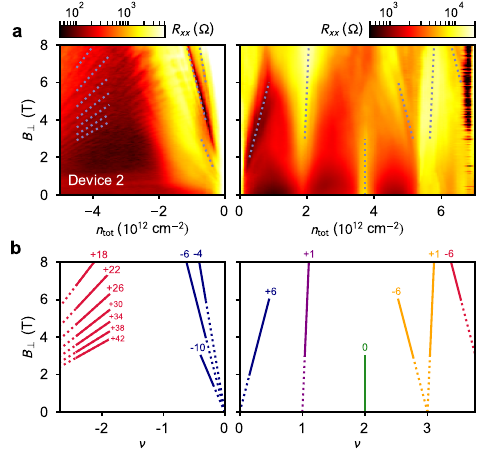}
    \caption{\figtitle{Twist angle determination.}
    \panel{a}, $R_{xx}$ Landau fan from Device~2, measured at $D=0$ and $T=\SI{300}{mK}$.
    Electron-side (right) and hole-side (left) are plotted with different color scales to improve contrast.
    Dashed lines correspond to the best-fit series shown in (\panel{b}).
    \panel{b}, Map of the best fit slopes from (\panel{a}) emerging from a consistent set of integer fillings, $\nu$.
    Red lines emerge from $n_{\nu=\pm 4}$ and $\nu=\pm 4$ (off-scale due to measurement limitations) in (\panel{a}) and (\panel{b}), respectively.
    }
    \label{fig:fanfit}
\end{figure}

We calibrate the twist angles for our HTG devices using the densities from which Landau levels emerge at integer fillings, particularly from $|\nu|=4$ (\extlbl~\ref{fig:fanfit}).
We fit a series of integer slopes to the measured Landau level gaps (dips in $R_{xx}$ emerging from $\nu=-4,0,+4$) and resistive states at partial fillings ($R_{xx}$ peaks at $\nu=1,2,3$), using the density of $|\nu|=4$, $n_{\nu=\pm 4}$, as a free parameter.
The best fit across all fillings and sloped features yields for device 2 $n_{\nu=\pm 4} = \SI{7.45 \pm 0.17e12}{cm^{-2}}$, corresponding to $\theta = \SI{1.79 \pm 0.02}{\degree}$.
Errors are estimated by aligning the collection of $R_{xx}$ features to the left and right edges of each feature ($R_{xx}$ minima for Landau level gaps or peaks of correlated states).
See Table~\ref{table:devices} for the twist angles and error estimates of the other devices (errors estimated from $R_{xx}$ peaks at integer filling for cases without clear Landau levels), along with a summery of the filling fractions $\nu$ for which correlated $R_{xx}$ features and the AHE in $R_{yx}$ are observed.
The preponderance of correlated features and observations of the AHE clearly increase as the twist angle approaches \SI{1.79}{\degree}, though the precise behavior for angles larger than \SI{\sim 1.8}{\degree} remains to be explored in detail.
This evolution of correlated features is especially evident in measurements of $R_{xx}$ plotted versus $\nu$ for several devices in \extlbl~\ref{fig:otherangles}.

We note that in the case of a slight mismatch between the two twist angles, $\theta_{12} \neq \theta_{23}$, the system relaxes to a structure similar to the equi-angle one, only with a smaller \supermoire unit cell (see Methods~\ref{ssec:unequal-twist}).
Transport measurements only allow us to extract the resulting local twist angle $\theta$ in the periodic domains, which is between $\theta_{12}$ and $\theta_{23}$.

\subsection{HTG at other twist angles}\label{ssec:twistdependence}
\begin{figure}
    \centering
    \includegraphics[width=3.27in]{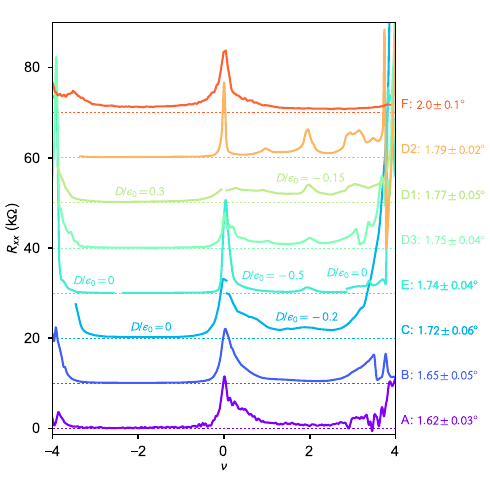}
    \caption{\figtitle{Twist angle dependence of $R_{xx}$ features.}
    Measured $R_{xx}$ traces from devices with a range of twist angles, as summarized in \exttbllbl~\ref{table:devices} (labels on the right correspond to entries in \exttbllbl~\ref{table:devices}).
    D1, D2, and D3 are equivalent to Devices 1, 2, and 3 shown in the paper, respectively.
    The traces were measured at a fixed $D$ field, with $D=0$ unless indicated otherwise (units are V/nm).
    The mean of the bounding estimates for the twist angle is used to label full filling, $\abs{\nu}=4$, for each curve.
    }
    \label{fig:otherangles}
\end{figure}

In addition to the three main devices reported here, we fabricated other equi-angle HTG devices at various twist angles ranging between \SI{1.6}{\degree} and \SI{2.0}{\degree}.
\extlbl~\ref{fig:otherangles} shows selected $R_{xx}(\nu)$ traces at constant $D$-fields, as indicated.
It reveals the range of twist angles that support correlated phases that appear as $R_{xx}$ peaks near integer $\nu$.
The results are summarized in Table~\ref{table:devices}.
We find correlated states in the range $\SI{1.7}{\degree} \lesssim \theta \lesssim \SI{1.8}{\degree}$ with the $\nu=2$ correlated phase surviving to the lowest twist angle.

\subsection{Symmetrization and antisymmetrization}\label{ssec:symmetrization}
All the presented $R_{xx}$ and $R_{yx}$ data were symmetrized and antisymmetrized, respectively, with respect to the applied out-of-plane magnetic field $B$.
Specifically, $R_{xx}=(R_{xx}^\mathrm{raw}(B)+R_{xx}^\mathrm{raw}(-B))/2$ and $R_{yx}=(R_{yx}^\mathrm{raw}(B)-R_{yx}^\mathrm{raw}(-B))/2$, where $\mathrm{raw}$ indicates raw data.
This allows us to compensate for non-ideal Hall bar geometry and for anisotropies that we found to be significant and ubiquitous in HTG.
In measurements where $B$ is the fast sweep axis, such as in Figs.~\ref{fig:AHE}\panel{a-f}, the symmetrization and antisymmetrization were performed between curves of opposite sweep direction, so that $R_{yx}=(R_{yx}^{raw \uparrow}(B)-R_{yx}^\mathrm{raw \downarrow}(-B))/2$ and $R_{xx}=(R_{xx}^\mathrm{raw \uparrow}(B)+R_{xx}^\mathrm{raw \downarrow}(-B))/2$ (here, the arrows indicate the sweep direction of $B$).
In measurements where $B$ was the slow axis, such as Figs.~\ref{fig:mosaic}\panel{a,b}, the antisymmetrization was performed between curves with opposite constant $B$: $R_{yx}=(R_{yx}^\mathrm{raw}(B)-R_{yx}^\mathrm{raw}(-B))/2$ and similarly for $R_{xx}$.

\subsection{Calculation of the Hall density}\label{ssec:Halldensity}
We extract the Hall density $n_\mathrm{H}=-e^{-1} (dR_{yx}/dB)^{-1}$ from $R_{yx}$ measurements taken at $\pm\SI{1}{T}$ according to $n_\mathrm{H}=-(B_{+}-B_{-})/e(R_{yx}(B_{+})-R_{yx}(B_{-}))$, where $B_{\pm}=\pm\SI{1}{T}$, $e$ is the elementary charge, and $n_\mathrm{H}>0$ corresponds to electron doping.
In a noninteracting system at low doping $n_\mathrm{H} \approx n$, and is expected to diverge near a VHS.
This is indeed the case for hole doping in our system.
In contrast, on the electron doping, near $\nu=1$ we find that $n_{H}$ deviates below $n$, indicative of a flavor reset \cite{Zondiner2020cascade,Wong2020cascade}, similar to magic-angle TBG (MATBG) and related \moire systems.

\subsection{Extraction of Curie temperature}\label{ssec:Curie}
\begin{figure*}
    \centering
    \includegraphics[width=3.27in]{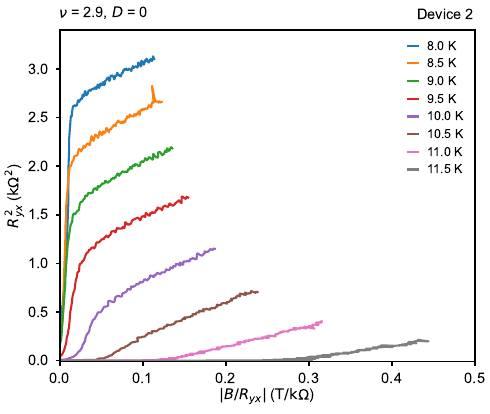}
    \caption{\figtitle{Extraction of the Curie temperature using an Arrott plot.}
    $R_{yx}^2$ versus $|B/R_{yx}|$.
    positive (negative) extrapolated intercept of the linear part at high $B$ indicates a ferromagnetic (paramagnetic) state.
    The curve taken at $T=\SI{10.5}{K}$ has approximately zero intercept, indicating a Curie temperature $T_\mathrm{C}\approx\SI{10.5}{K}$}.
    \label{fig:Arrott}
\end{figure*}

We extract $T_\mathrm{C}$ following Ref.~\cite{Chiba2011Electrical} by plotting $R_{yx}^2$ versus $|B/R_{yx}|$ at different temperatures.
Such a plot for Device 2 is shown in \extlbl~\ref{fig:Arrott}, taken at $\nu=2.9$ and $D=0$.
We take $R_{yx}$ as a proxy for the magnetization $M$, reproducing an Arrott plot ( $M^2$ versus $M/H$) \cite{Arrott1957Criterion}.
The intercept of a linear extrapolation of the high-field regime determines the magnetic state.
At temperatures $T<\SI{10.5}{K}$ we find a positive intercept indicating a ferromagnetic phase.
At $T=\SI{10.5}{K}$ the intercept is approximately zero, and above this temperature, the intercept is negative indicating a transition to a paramagnetic phase, hence the Curie temperature is approximately \SI{10.5}{K}.

\subsection{Theoretical electronic band structure calculation}\label{ssec:bandstructure}
The theoretical single-particle band structure shown in \figlbl~\ref{fig:setup}\panel{e} was calculated using the continuum model Hamiltonian~\cite{devakul2023magicangle}
\begin{equation}
H = 
\begin{pmatrix}
v_0\vec{\sigma}_{\theta}\cdot [\vec{k}-\vec{K}_1] + U& T(\vec{r} -\vec{d}_t) & 0\\
T^\dag(\vec{r} -\vec{d}_t)& v_0\vec{\sigma} \cdot [\vec{k}-\vec{K}_2] &  T(\vec{r}-\vec{d}_b) \\
0 & T^\dag(\vec{r}-\vec{d}_b) & v_0\vec{\sigma}_{-\theta}\cdot [\vec{k}-\vec{K}_3]-U\\
\end{pmatrix}
\end{equation}\label{eq:contham}
where $(\vec{K}_1,\vec{K}_2,\vec{K}_3)=(k_\theta, 0, -k_\theta)\hat{y}$, with $k_\theta=\frac{8\pi}{3a_0}\sin(\theta/2)$ and $a_0=\SI{0.246}{nm}$, and $\vec{\sigma}_\theta =e^{-i\theta\sigma_z}(\sigma_x,\sigma_y)$ are Pauli matrices.
The interlayer tunneling terms are
\begin{equation}
T(\vec{r}) =
\begin{pmatrix}
w_{AA} U_0(\vec{r}) & w_{AB} U_{-1}(\vec{r}) \\
w_{AB} U_{1}(\vec{r}) & w_{AA} \kappa U_0(\vec{r}) 
\end{pmatrix}
\end{equation}
with $U_l(\vec{r})=e^{i\vec{q}_0\cdot\vec{r}}\sum_{n=0}^{2}e^{\frac{2\pi i n}{3} l n } e^{-i \vec{q}_n\cdot\vec{r}}$, $q_{n,x}+iq_{n,y}=-ik_\theta e^{\frac{2\pi i }{3}n}$.
The h-HTG and $\overline{h}$-HTG regions are modeled by choosing the displacements $\vec{d}_t-\vec{d}_b=\pm\vec{\delta}$ respectively, where $\delta=\frac{1}{3}(\vec{a}_2-\vec{a}_1)$, and $a_{1,2}=a_0(\pm\frac{\sqrt{3}}{2},\frac{1}{2})$ are the atomic lattice vectors.
The layer potential $U$ models the effect of the displacement field, up to electrostatic corrections.
We use parameters $v_0=8.8\times \SI{1e5}{m/s}$, $w_{AB}=\SI{110}{meV}$, and $w_{AA}=\SI{75}{meV}$, and typically neglect the Pauli matrix rotation $\sigma_{\theta}\rightarrow\sigma$.

\subsection{Microscopic parameters and valence band Van Hove singularity}\label{ssec:microscopic}
We use the shape of the VHS on the hole doping side to estimate the value of the effective velocity ratio $v_0/w_{AB}$.  
Since the hole side ($n<0$) does not show strong interaction effects in experiment, we can approximately treat the valence bands as non-interacting but with a renormalized velocity $\tilde{v}_0$.  
The shape of the observed VHS can then be used to constrain the value of the $\tilde{v}_0/w_{AB}$.

\begin{figure}
    \centering
    \includegraphics[height=3.5in]{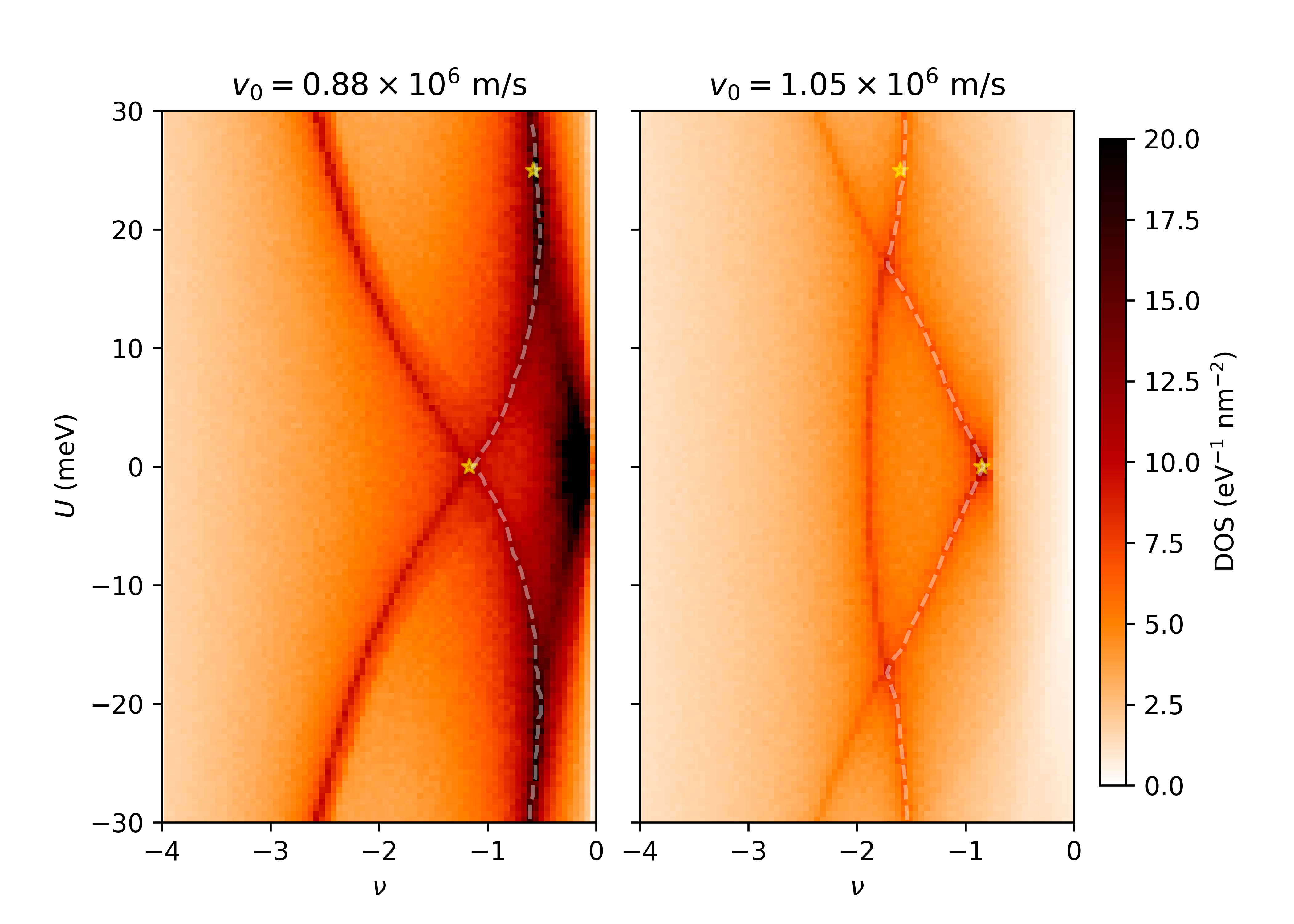}
    \includegraphics[height=3.5in]{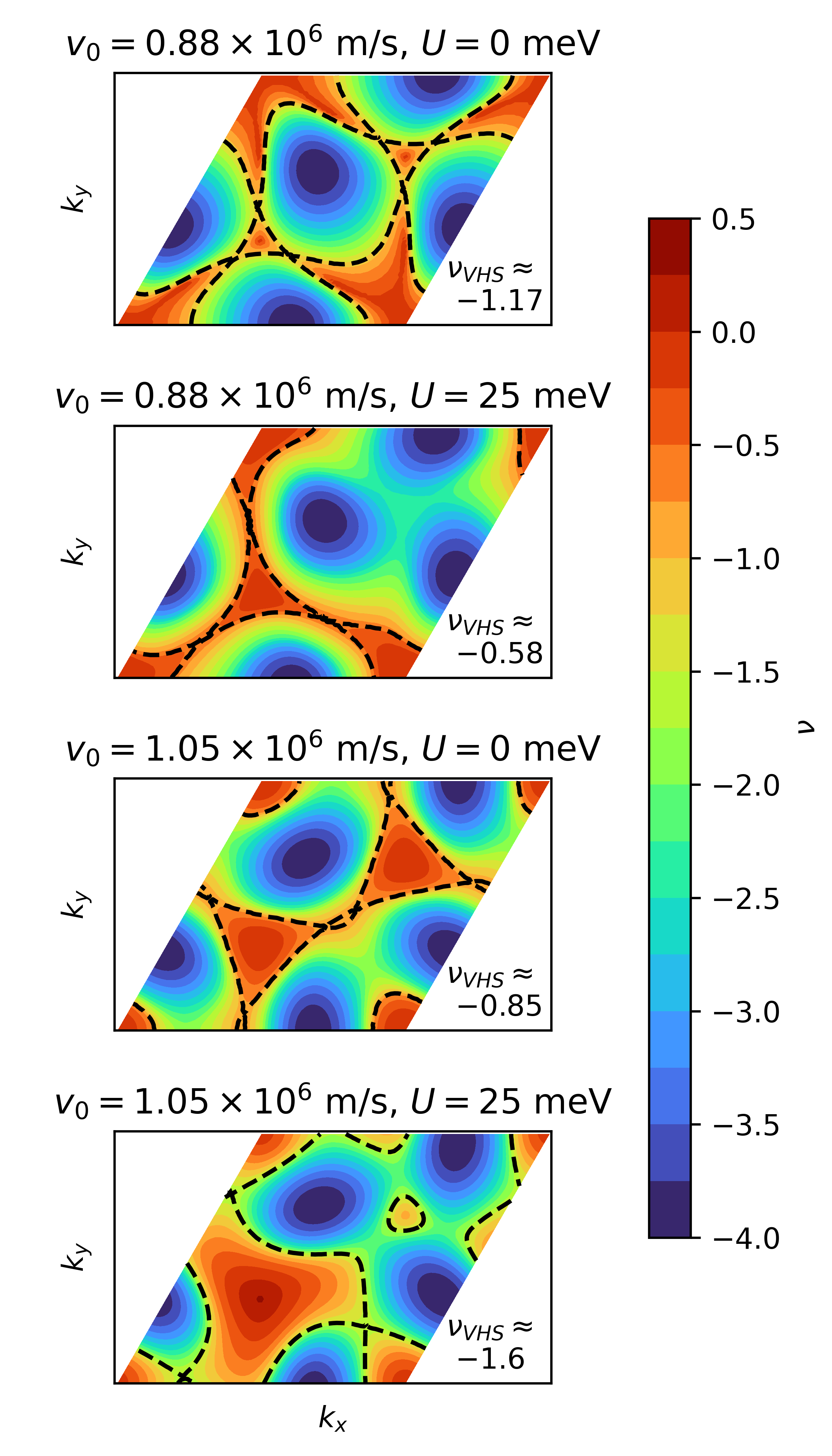}
    \caption{\figtitle{Single-particle density of states and Van Hove singularity.}
    (left and center) The single particle DOS for hole doping as a function of filling factor and layer potential, for two velocity parameters $v_0$.
    The VHS at which the Hall density switches sign is identified by the dashed lines.  
    (right) Extended Fermi surfaces at the VHS are shown for the four points indicated by stars in the DOS plot.
    }
    \label{fig:vhs}
\end{figure}

\extlbl~\ref{fig:vhs}(left) shows the non-interacting density of states (DOS) calculated for the continuum model Eq~\ref{eq:contham} as a function of filling factors $-4<\nu<0$ and $U$, for two choices of velocity: a ``bare'' velocity $v_0=0.88\times \SI{1e6}{m/s}$ and a ``renormalized'' velocity $\tilde{v}_0=1.05\times \SI{1e6}{m/s}$.  
The DOS shows multiple peaks corresponding to VHS arising from various Lifshitz transitions in the Fermi surface.
To compare with \figlbl~\ref{fig:nD}\panel{b} of the main text, we identify the changing sign of the Hall density with the VHS at which extended orbits exist in the Fermi surface \cite{ashcroft2022solid}.
\extlbl~\ref{fig:vhs}(right) shows the Fermi surfaces of the valence band as a function of filling factor at $U=0$ and $U=\SI{25}{meV}$, which allow for the identification of the VHS with extended orbits, indicated by stars and the dashed lines in \extlbl~\ref{fig:vhs}(left).

The observed VHS for hole doping in \figlbl~\ref{fig:nD}\panel{b}, which appears at $\nu\approx -0.8$ for $U=0$, and moves to higher hole doping $\nu<-0.8$ in a displacement field, is therefore in better agreement with the theoretical model using an effective renormalized velocity $\tilde{v}_0$.

\subsection{Hartree-Fock calculations at $\nu=7/2,2/3$}
\label{ssec:HF}

In this section, we perform self-consistent Hartree-Fock calculations in moir\'e-periodic h-HTG at non-integer fillings $\nu=7/2,2/3$, the same fillings where correlated features were observed in \figlbl~\ref{fig:AHE}.  To the non-interacting Hamiltonian in Section~\ref{ssec:bandstructure}, we apply layer potentials $U,0,-U$ on the three layers to mimic the effect of an external displacement field. Owing to the large energy gap to the remote bands, we project our calculations into the two central bands per flavour (spin and valley). We add dual-gate screened density-density interactions $V(q)=\frac{e^2}{2\epsilon_0\epsilon_r q}\tanh qd_{\text{sc}}$, where the gate screening length is $d_{\text{sc}}=25\,\text{nm}$, and the effect of the hBN dielectric and remote bands is phenomenologically captured with the relative permittivity $\epsilon_r=8$.  The interaction term is normal-ordered with respect to the average density of the central bands at charge neutrality. To allow for gapped states at non-integer fillings within mean-field theory, we allow translation symmetry-breaking (TSB) by enlarging the unit cell. We allow breaking of all flavor and discrete rotational symmetries. Further details of the Hartree-Fock procedure are provided in Ref.~\cite{kwan2023strongcoupling}.

\begin{figure}
    \centering
    \includegraphics[width=6in]{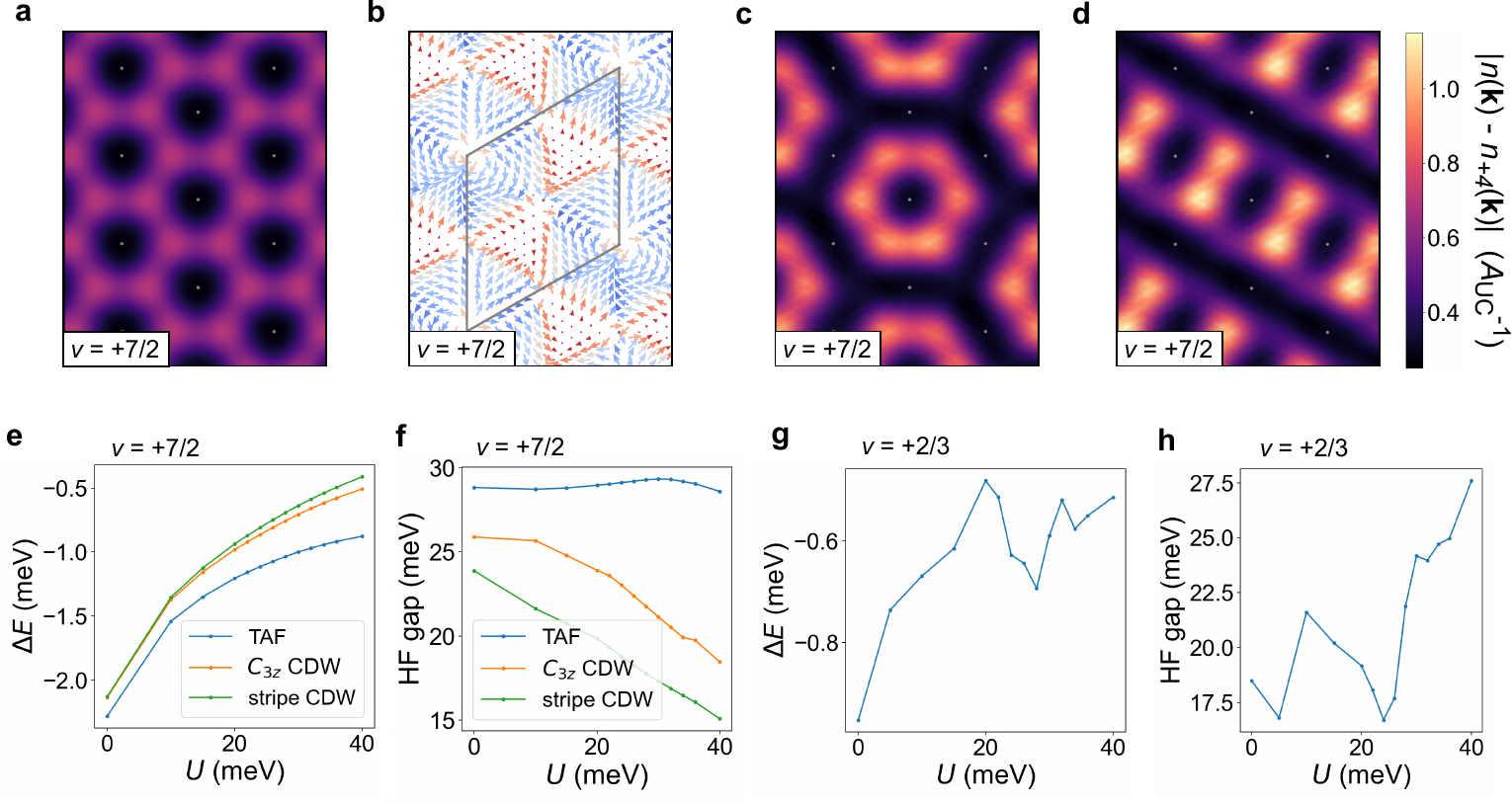}
    \caption{\figtitle{Hartree-Fock calculations at $\nu=7/2$ and $2/3$. }
    \textbf{a}, Charge density $n(\bm{r})$ (measured relative to that at full flat band filling $n_{+4}(\bm{r})$) of the tetrahedral antiferromagnet (TAF) at $\nu=7/2$. Grey dots indicate ABA-stacking regions. \textbf{b}, Local spin orientation in the TAF. Arrows denote spin direction in $s_x-s_y$ plane, while red (blue) coloring indicates out-of-plane polarization along $+\hat{s}_z$ ($-\hat{s}_z$). Grey parallelogram indicates the new quadrupoled moir\'e unit cell. \textbf{c,d} Same as \textbf{a} except for the $\hat{C}_{3z}$ CDW and stripe CDW respectively. \textbf{e},
    $\Delta E$ of the different translation symmetry breaking solutions at $\nu=7/2$ as a function of interlayer potential $U$. $\Delta E$ is measured relative to that of the best translation-symmetric solution. 
    \textbf{f}, Charge gap of the translation symmetry breaking solutions at $\nu=7/2$. \textbf{g,h}, Same as \textbf{e,f} except for the best translation symmetry breaking solution at $\nu=2/3$. All calculations performed on a $18\times18$ system using $\theta=1.80^\circ,w_{AA}=75\,\text{meV}$.
    }
    \label{fig:HF}
\end{figure}

At $\nu=7/2$ (\extlbl~\ref{fig:HF}\panel{a}-\panel{f}), we additionally let the system break translation symmetry by doubling the unit cell length along both moir\'e axes (quadrupling the area of the unit cell). We find that the lowest energy solution is a gapped state with TSB, as shown by the negative value of $\Delta E$, which is defined as the Hartree-Fock energy per moir\'e unit cell, measured relative to the best translation-symmetric solution. Its density matrix is consistent with fully filling all bands, except a $|C|=2$ band (corresponding to a $B$ sublattice band in h-HTG) which is half-filled and reconstructed by TSB and spatially-dependent spin rotations. While the charge density is moir\'e-periodic (\extlbl~\ref{fig:HF}\panel{a}), the quadrupling of the unit cell is revealed by the non-coplanar spin texture (\extlbl~\ref{fig:HF}\panel{b}) which forms a tetrahedral antiferromagnet (AFM), similar to that theoretically proposed in twisted monolayer-bilayer graphene (TMBG) and twisted double bilayer graphene (TDBG) in Ref.~\cite{Wilhelm2023noncoplanar}. If we restrict the calculation to maintain spin-collinearity, we find two other spin-polarized solutions that realize a $C_{3z}$-symmetric CDW and a stripe CDW respectively (\extlbl~\ref{fig:HF}\panel{c}-\panel{d}). The stripe CDW is reminiscent to that proposed in Ref.~\cite{Polshyn2022Topological} to explain transport experiments in TMBG. All three solutions are $|C|=1$ states that preserve valley $U(1)_V$ symmetry, and closely resemble the candidate `strong-coupling' TSB orders expected from half-filling an ideal $|C|=2$ band~\cite{Wilhelm2023noncoplanar,Dong2023manybody}. The close energetic competition between the different orders (\extlbl~\ref{fig:HF}\panel{e}) points towards the ideality of the topological bands in h-HTG. $|\Delta E|$ decreases monotonically as a function of $U$, suggesting that the TSB state is weakened in a displacement field, though the $U$-dependence of the charge gap is less consistent (\extlbl~\ref{fig:HF}\panel{f}). While the TSB solutions remain energetically favoured for the large range of interlayer potentials studied, we caution that Hartree-Fock tends to overestimate gaps and symmetry-breaking, such that beyond mean-field theory, the threshold value of $U$ where the system recovers symmetry is expected to be reduced. 

In \extlbl~\ref{fig:HF}\panel{g}-\panel{h}, we show analogous results for $\nu=2/3$, where we allow translation symmetry-breaking to enlarge the unit cell threefold along both moir\'e axes. We again find the presence of gapped TSB solutions, though the gap size and TSB energy gain $|\Delta E|$ are smaller and non-monotonic in $U$. Interestingly, we find a window of non-zero interlayer potentials, slightly above/at the theoretical topological transition for $\nu=1$~\cite{kwan2023strongcoupling}, where $|\Delta E|$ is locally maximal and the HF gap remains large. This suggests the possibility of a correlated state that only emerges at a nonzero displacement field. However, the presence of several closely-competing states, multiple partially filled flavors, and sensitive dependence on system parameters prevent an unambiguous interpretation of the Hartree-Fock results. We leave a more detailed theoretical investigation of the correlated physics at $\nu=2/3$ to future work.

\subsection{Magnetization reversal}
\label{ssec:magnetizationreversal}
In this section, we show using a theoretical model that the orbital magnetization of h-HTG switches sign as the density is tuned across the correlated insulating gap at $\nu=3$.
This explains the density-induced magnetization switching observed in \figlbl~\ref{fig:mosaic}, similar to the Chern mosaic in hBN-aligned MATBG~\cite{Grover2022}

\begin{figure}
    \centering
    \includegraphics[height=2.3in]{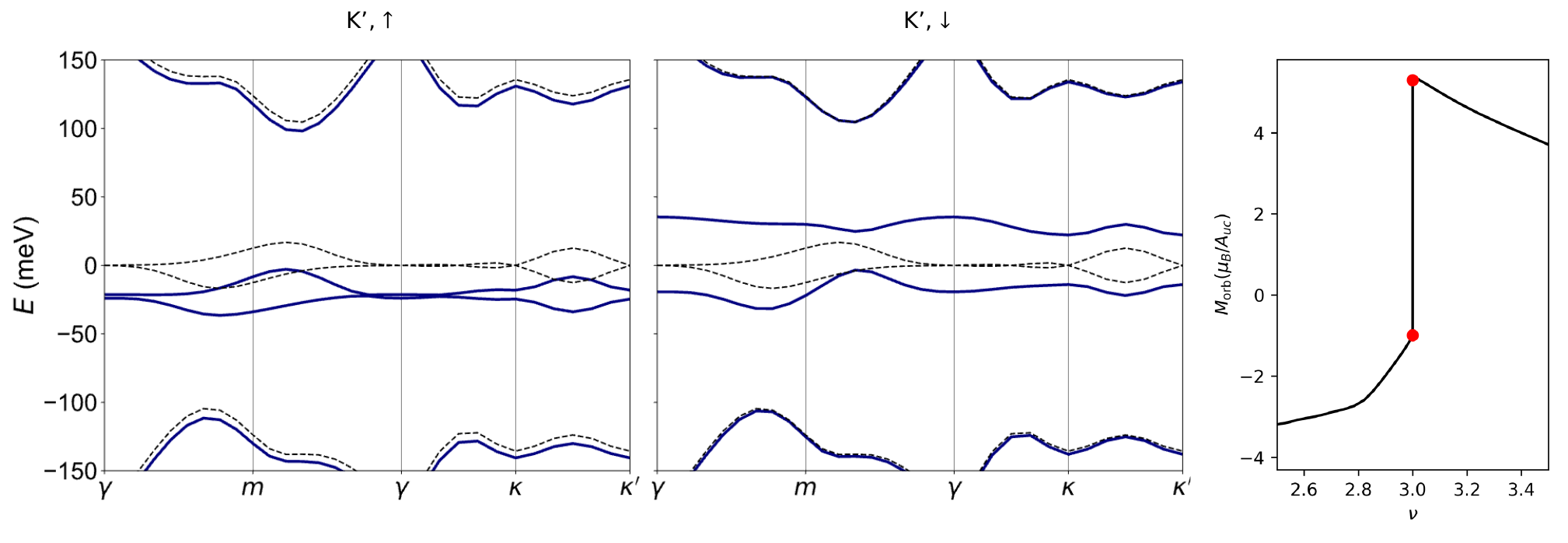}
    \caption{\figtitle{Hartree-Fock bands and orbital magnetization at $\nu=3$.}
    (lefta and center) The HF quasiparticle bands at $\nu=3$ are shown for the $(K^\prime,\uparrow)$ (left) and $(K^\prime,\downarrow)$ (center) sectors, with the non-interacting bands shown in dashed lines.  
    The $K$ valley bands are  identical to the $(K^\prime,\uparrow)$, up to $k\rightarrow -k$.
    (right) The orbital magnetization, in units of Bohr magneton per moir\'e unit cell, calculated for the simplified model, showing a discontinuity and sign flip at $\nu=3$ (red dots).
       }
    \label{fig:M_orb}
\end{figure}

In \extlbl~\ref{fig:M_orb}(left), we show the self-consistent Hartree-Fock band structure at $\nu=3$.
The self-consistent calculation is performed including the remote bands.
The single-particle bands are shown in the dashed lines.
We observe that each of the filled bands is (quite rigidly) shifted, with the filled bands being shifted down by $\Delta_-\approx \SI{20}{meV}$, and the unfilled band being shifted up by $\Delta_+\approx \SI{30}{meV}$, while the remote bands remain mostly unchanged \cite{kwan2023strongcoupling}.
In this case, the unfilled band is the (valley,spin,Chern-sublattice)=$(K^\prime,\downarrow,A)$ band.

To demonstrate the switching of orbital magnetization, we consider a simplified model following the Hartree-Fock bands.
We assume the remote bands are perfectly rigid, and model the interaction-induced band shifting by a term
$\Delta_- P + \Delta_+ (1-P) $,
where $P$ is a (momentum-dependent) projector to the filled Chern-sublattice-basis bands in the flat band manifold,
which rigidly shifts the filled bands by $\Delta_-=\SI{-20}{meV}$ and unfilled bands by $\Delta_+=\SI{30}{meV}$.  
The orbital magnetization can then be computed as a function of the chemical potential $\mu$ (which can be arbitrary, but is only realistic if it also corresponds to the physical density $\nu=3$), given by Ref.~\cite{Xiao2010Berry}
\begin{equation}
\begin{split}
M_{\mathrm{orb}}=-\frac{e}{\hbar}\sum_n\int \frac{d^2k}{4\pi^2} f(E_n(k)-\mu) \sum_{m\neq n}\mathrm{Im}\frac{\bra{n(k)}\partial_{k_x}H\ket{m(k)}\bra{m(k)}\partial_{k_y}H\ket{n(k)}}{(E_n(k)-E_m(k))^2} \\
\times \left[(E_n(k)-E_m(k))+2(\mu-E_n(k))\right]
\end{split}
\end{equation}
where $\ket{n(k)}$ are the Bloch wavefunctions at $k$, $n,m$ vary over all band indices (not just those in the flat band manifold), $E_n(k)$ is the energy of the $n$th band, and $f(x)=(1+e^{\beta x})^{-1}$ is the Fermi function.
Because $P$ leaves the remote bands unchanged, the contribution to the magnetization from bands far from the Fermi energy vanishes.
Equivalently, this can be written
\begin{equation}
M_{\mathrm{orb}}=-\frac{e}{\hbar}\sum_n\int \frac{d^2k}{4\pi^2} f(E_n(k)-\mu) \sum_{m\neq n}\mathrm{Im}
\bra{\partial_{k_x} n(k)} (E_n(k)-H)+2(\mu-E_n(k))\ket{\partial_{k_y} n(k)}
\end{equation}
Importantly, $M_{\mathrm{orb}}$ depends on $\mu$ even in the gap, due to contributions from the edge modes to the orbital magnetization.
This allows $M_{\mathrm{orb}}$ to be discontinuous and flip sign when the density is tuned from just below to just above $\nu=3$.

In \extlbl~\ref{fig:M_orb}(right), we show the orbital magnetization as a function of the filling factor $\nu$.  We use a $60\times 60$ discretization of the moir\'e Brillouin zone and $\beta^{-1}=\SI{0.2}{meV}$.
The orbital magnetization is discontinuous and indeed switches sign from negative to positive at $\nu=3$.
In addition to the state studied here, there is also the time-reversed partner which has opposite $M_{\mathrm{orb}}$.
The energetic competition between these two states in an applied magnetic field can therefore be affected by tuning density to directly above/below the $\nu=3$ gap.

\subsection{Relaxation}
\label{ssec:relaxation}
To calculate the relaxation of the HTG system, we employ a continuum relaxation model in local configuration space~\cite{zhu2020relaxation}.
Therefore, instead of formulating the problem in real space, we adopt configuration space, which describes the local environment of every position in layer $L_\ell$ and bypasses a periodic approximation~\cite{cazeaux2019}. 
Every position in real space $\vec{r}$ in $L_i$ can be uniquely parametrized by three shift vectors $\vec{b}^{i\rightarrow j}$ for $j = 1, 2, 3$ that describes the relative position between any point in real space $\vec{r}$ with respect to all three layers. Note that $\vec{b}^{i \rightarrow j} = \vec{0}$ if $i = j$ since the separation between a position with itself is 0, which leads to a four-dimensional configuration space. 

For a given real space position $\vec{r}$, the following linear transformation uniquely maps between the real space position, $\vec{r}$, and the local configuration space component in layer $i$ with respect to layer $j$ $\vec{b}^{i\rightarrow j}$:
\begin{equation}
\vec b^{i\rightarrow j} (\vec{r}) = (E_j^{-1} E_i - \mathbbm{1}) \vec r, \label{eqn:mapping}
\end{equation}
where $E_i$ and $E_j$ are the unit cell vectors of layers $i$ and $j$ respectively, rotated by $\theta_{ij}$. In the trilayer system, there is no simple linear transformation between real and configuration space. The relation between the displacement field defined in real space, $\vec{U}^{(i)} (\vec{r})$, and in configuration space, $\vec{u}^{(i)} (\vec{b}) $, can be found by evaluating $\vec{u}^{(j)} (\vec{b}) $ at the corresponding $\vec{b}^{i\rightarrow j} (\vec{r})$ and $\vec{b}^{i\rightarrow k} (\vec{r}) $ with Eq.~\eqref{eqn:mapping} to obtain 
\begin{equation}
    \vec{U}^{(i)} (\vec{r}) = \vec{u}^{(i)}(\vec{b}^{i\rightarrow j} (\vec{r}), \vec{b}^{i\rightarrow k} (\vec{r})), 
\end{equation}
where $j, k \neq i$ and $j < k$.

The relaxed energy has two contributions, intralayer and interlayer energies:
\begin{align}
	&E^\rr{tot} (\bu^{(1)}, \bu^{(2)}, \bu^{(3)}) = E^\rr{intra} (\bu^{(1)}, \bu^{(2)}, \bu^{(3)}) + E^\rr{inter} (\bu^{(1)}, \bu^{(2)}, \bu^{(3)}), \label{eq:total_E}
\end{align}
where $\bu^{(\ell)}$ is the relaxation displacement vector in layer $\ell$. 
To obtain the relaxation pattern, we minimize the total energy with respect to the relaxation displacement vector. 

We model the intralayer coupling based on linear elasticity theory:
\begin{align}
E^\mathrm{intra} (\vec u^{(1)}, \vec u^{(2)}, \vec u^{(3)}) 
&= \sum_{\ell=1}^3 \int \frac{1}{2} \Big[G (\partial_x u^{(\ell)}_x + \partial_y u^{(\ell)}_y)^2  \nonumber \\
&\ \ \  + K ( (\partial_x u^{(\ell)}_x - \partial_y u^{(\ell)}_y)^2 + (\partial _x u^{(\ell)}_y + \partial_y u^{(\ell)}_x)^2) \Big] d \vec{b},\label{eqn:intra}
\end{align}
where $G$ and $K$ are shear and bulk moduli of monolayer graphene, which we take to be $G = 47352 \, \mathrm{meV/unit \ cell}$, $K = 69518 \, \mathrm{meV/unit \ cell}$~\cite{carr2018relaxation,zhu2020relaxation}.  

The interlayer energy accounts for the energy cost of the layer misfit, which is described by the generalized stacking fault energy (GSFE) ~\cite{Kaxiras1993,Zhou2015}, obtained using first principles Density Functional Theory (DFT) with the Vienna Ab initio Simulation Package (VASP)~\cite{Kresse1993, Kresse1996a, Kresse1996b}. GSFE is the ground state energy as a function of the local stacking with respect to the lowest energy stacking between a bilayer. For bilayer graphene, GSFE is maximized at the AA stacking and minimized at the AB stacking. Letting $\vec{b} = (b_x, b_y)$ be the relative stacking between two layers, we define the following vector $\bm v= (v, w) \in [0, 2 \pi ]^2$:
\begin{equation}
	\begin{pmatrix}
		v \\ w
	\end{pmatrix}
		= \frac{2 \pi}{a_0} \mqty[\sqrt{3}/2 & -1/2 \\ \sqrt{3}/2 & 1/2]
	\begin{pmatrix}
		b_x \\ b_y
	\end{pmatrix},
\end{equation}
where $a_0 = \SI{2.4595}{\angstrom}$ is the graphene lattice constant. 
We parameterize the GSFE as follows, 
\begin{align}
V^\mathrm{GSFE}_{j\pm} = c_0 + & c_1(\cos v + \cos w + \cos (v + w) ) \nonumber \\
+ & c_2 (\cos (v + 2w) + \cos(v-w) + \cos(2 v + w)) \nonumber \\
+ & c_3 (\cos(2 v ) + \cos (2 w) + \cos(2 v + 2 w)),\label{eqn:vgsfe}
\end{align}
where we take $c_0 = 6.832\, \mathrm{meV/cell}$, $c_1 = 4.064 \, \mathrm{meV/cell}$, $c_2 = -0.374 \, \mathrm{meV/cell}$, $c_3 = -0.0095\, \mathrm{meV/cell}$~\cite{zhu2020relaxation,carr2018relaxation}.
The van der Waals force is implemented through the vdW-DFT method using the SCAN+rVV10 functional~\cite{Peng2016}. In terms of $V^\mathrm{GSFE}_{\ell\pm}$, the total interlayer energy can be expressed as follows:
\begin{align}
	E^\mathrm{inter} &= \frac{1}{2}\int \misfitr{1+}{1}{2}\,\mathrm{d} \vec{b} +\frac{1}{2}\int \left[ \misfitr{2-}{2}{1} + \misfitr{2+}{2}{3} \right]\,\mathrm{d}\vec{b} \nonumber \\
&+\frac{1}{2}\int \misfitr{3-}{3}{2} \,\mathrm{d}\vec{b}, \nonumber
\end{align}
where $\vec{B}^{i\rightarrow j} = \vec{b}^{i\rightarrow j} + \bu^{(j)} - \bu^{(i)}$ is the relaxation modified local shift vector.
Note that we neglect the interlayer coupling between layers 1 and 3. 
The total energy is obtained by summing over uniformly sampled configuration space. In this work, we discretize the four-dimensional configuration space by $54 \times 54 \times 54 \times 54$.

\subsection{Unequal twist angles}
\label{ssec:unequal-twist}
\begin{figure}
    \centering
    \includegraphics[height=2.8in]{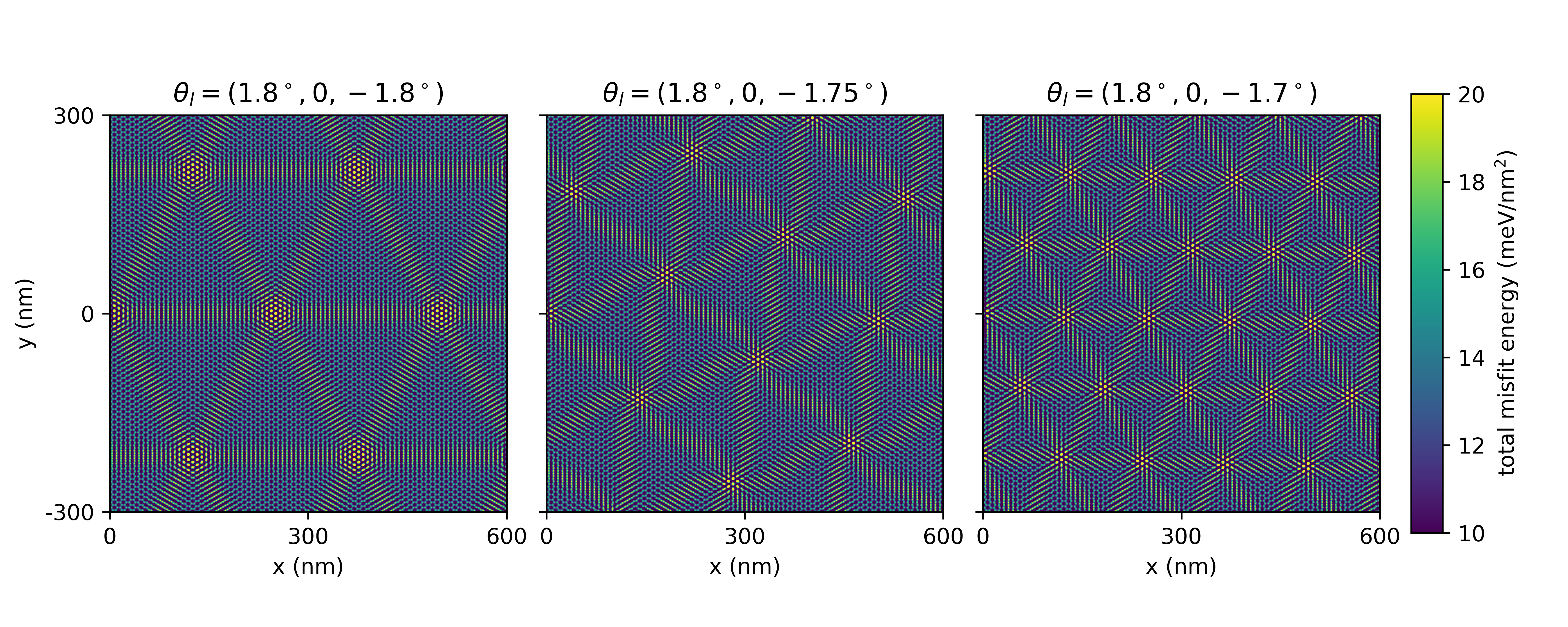}
    \caption{\figtitle{Relaxed structure for unequal twist angles.}
    The interlayer misfit energy, summed over both layer pairs, is shown for HTG with angles $\theta_l=(\theta,0,-\theta^\prime)$.
    The main effect of a small angle mismatch $\theta\neq\theta^\prime$ is a reduction in the supermoir\'e domain size.
    }
    \label{fig:unequalangles}
\end{figure}

We show the relaxed supermoir\'e structure calculated for unequal twist angles $(\theta,0,-\theta^\prime)$.
\extlbl~\ref{fig:unequalangles} shows the local misfit energy for $\theta=1.8^\circ$ with varying $\theta^\prime=1.8^\circ,1.75^\circ,1.7^\circ$.  
The h-HTG and \hb-HTG domains can be identified by the honeycomb pattern in the misfit energy.
It can be seen that the local physics within the h-HTG and \hb-HTG domains remain relatively unchanged, meaning that the structure locally relaxes into the equal-angle commensurate configuration.
Thus, the main effect of the angle difference is the reduction of the domain size (which is determined by the supermoir\'e period of the unrelaxed structure).

\acknowledgments
We thank M. Kastner and P. Ledwith for helpful discussions, A. Bangura, G. Jones, R. Nowell, A. Woods, and S. Hannahs for technical support, and X. Wang for assistance with device fabrication.
This work was partially supported by the Army Research Office MURI W911NF2120147, the 2DMAGIC MURI FA9550-19-1-0390, the National Science Foundation (DMR-1809802), the STC Center for Integrated Quantum Materials (NSF grant no. DMR-1231319), and the Gordon and Betty Moore Foundation’s EPiQS Initiative through grant GBMF9463 to PJH. 
This work was supported by the Air Force Office of Scientific Research (AFOSR) under award FA9550-22-1-0432.
Measurement infrastructure was funded in part by the Gordon and Betty Moore Foundation’s EPiQS initiative through grant GBMF3429 and grant GBMF9460. 
D.G.-G. gratefully acknowledges support from the Ross M. Brown Family Foundation.
K.W. and T.T. acknowledge support from the JSPS KAKENHI (Grant Numbers 21H05233 and 23H02052) and World Premier International Research Center Initiative (WPI), MEXT, Japan.
A portion of this work was performed at the National High Magnetic Field Laboratory, which is supported by National Science Foundation Cooperative Agreement No. DMR-2128556 and the State of Florida.
This work was performed in part at the Harvard University Center for Nanoscale Systems (CNS); a member of the National Nanotechnology Coordinated Infrastructure Network (NNCI), which is supported by the National Science Foundation under NSF award no. ECCS-2025158.
This work was carried out in part through the use of MIT.nano's facilities.
This work made use of the MRSEC Shared Experimental Facilities at MIT, supported by the National Science Foundation under award number DMR-1419807.
AU acknowledges support from the MIT Pappalardo Fellowship and from the VATAT Outstanding Postdoctoral Fellowship in Quantum Science and Technology.
ZZ is supported by a Stanford Science fellowship. 
Sandia National Laboratories is a multimission laboratory managed and operated by National Technology \& Engineering Solutions of Sandia, LLC, a wholly owned subsidiary of Honeywell International Inc., for the U.S. Department of Energy’s National Nuclear Security Administration under contract DE-NA000352.

\section*{Author contributions}

S.C.d.l.B., and A.U. conceived the project. L.-Q.X. fabricated the devices with the help of A.U.. L.-Q.X., A.U. and S.C.d.l.B. carried out the helium-3 transport measurements. A.S. and L.-Q.X. carried out the dilution fridge transport measurements under the supervision of D.G.-G.. T.D. and Y.H.K. performed band structure, magnetization and Hartree-Fock calculations. T.D. and Z.Z performed lattice relaxation calculations. K.W. and T.T. supplied the boron nitride crystals. A.U., S.C.d.l.B., L.-Q.X., A.S., T.D., L.F., and P.J-H. analyzed the data and discussed the interpretation. A.U., S.C.d.l.B., and L.-Q.X. wrote the manuscript with input from all authors. P.J.-H. supervised the project.

\section*{Competing interests}

The authors declare no competing interests.

\section*{Data availability}

The data that support the findings of this study are available from the corresponding authors upon reasonable request.

\bibliography{ref}

\clearpage
\ExtendedData

\begin{table*}
    \begin{booktabs}{ccc}
    \toprule
    \textbf{System} & $T_\mathrm{C}$ (K) & \textbf{Reference} \\
    \midrule[solid]
    \SetCell[r=4]{c} Twisted bilayer MoTe$_2$ & 14 & \cite{anderson2023programming,Cai2023Signatures} \\ 
    & 13 & \cite{zeng2023integer} \\
    & 12 & \cite{park2023observation} \\
    & 10 & \cite{xu2023observation} \\
    \midrule[dotted]
    HTG & 10.5 & This work \\
    \midrule[dotted]
    \SetCell[r=2]{c} hBN-aligned MATBG & 7.5 & \cite{Serlin2020Intrinsic} \\ 
    & 5 & \cite{Sharpe2019Emergent} \\
    \midrule[dotted]
    \SetCell[r=2]{c} tMBG & 7 & \cite{Polshyn2020electrical} \\ 
    & 2.5 & \cite{Chen2021Electrically} \\
    \midrule[dotted]
    \SetCell[r=2]{c}
    WSe$_2$/MATBG & 7 & \cite{Lin2022spinorbit} \\
    & 5 & \cite{polski2022hierarchy} \\
    \midrule[dotted]
    AB-stacked MoTe$_2$/WSe$_2$ & 5.5 & \cite{li2021quantum} \\
    \midrule[dotted]
    Non-magic-angle twisted bilayer graphene & 5.5 & \cite{Tseng2022anomalous} \\
    \midrule[dotted]
    Near-commensurate hBN-MATBG superlattice & 4.5 & \cite{Stepanov2021Competing,Grover2022} \\
    \midrule[dotted]
    hBN-aligned rhombohedral trilayer graphene & 3.5 & \cite{Chen2020Tunable} \\
    \midrule[dotted]
    WSe$_2$/magic-angle twisted trilayer graphene & 2.5 & \cite{zhang2023valley} \\
    \midrule[dotted]
    AB-AB stacked twisted double bilayer graphene & 2.5 & \cite{Kuiri2022} \\
    \midrule[dotted]
    AB-BA stacked twisted double bilayer graphene & $<$2 & \cite{he2021chiralitydependent} \\
    \bottomrule
    \end{booktabs}
    \caption{\figtitle{Summary of AHE reported in \moire systems to date.}
    }
    \label{table:Tc}
\end{table*}

\sisetup{range-phrase=--}
\begin{table*}
    \begin{booktabs}{ccccc}
    \toprule
    \textbf{Label} & \textbf{Extracted angle}    & \textbf{$\nu$, correlated features}   & \textbf{$\nu$, AHE}  & \textbf{$e$-$h$ symmetry} \\
    \midrule[solid]
    A & \SI{1.62\pm 0.03}{\degree}  & - & - & Nearly symmetric \\ 
    \midrule[dotted]
    B & \SI{1.65\pm 0.05}{\degree}  & - & - & Nearly symmetric \\ 
    \midrule[dotted]
    C & \SI{1.72\pm 0.06}{\degree}  & 2 & - & Asymmetric \\ 
    \midrule[dotted]
    E & \SI{1.74\pm 0.04}{\degree}  & 2,3 & Unknown & Asymmetric \\ 
    \midrule[dotted]
    D3 & \SI{1.75\pm 0.04}{\degree}  & 1,2,3 & 1,3 & Asymmetric \\ 
    \midrule[dotted]
    D1 & \SI{1.77\pm 0.05}{\degree}  & 1,2,3 & 1,3 & Asymmetric \\ 
    \midrule[dotted]
    D2 & \SI{1.79\pm 0.02}{\degree}  & 1,2,3,$\frac{7}{2}$ & $\frac{2}{3}$,1,3 & Asymmetric \\ 
    \midrule[dotted]
    F & \SI{2.0\pm 0.1}{\degree} & - & - & Nearly symmetric \\ 
    \bottomrule
    \end{booktabs}
    \caption{\figtitle{Summary of twist angles with observed transport features.}
    }
    \label{table:devices}
\end{table*}

\end{document}